\documentclass{emulateapj}

\begin{document}

\shorttitle{Absorptive H$_\alpha$ Polarized Line Profiles}
\shortauthors{Harrington \& Kuhn}

\title{Ubiquitous H$_\alpha$ polarized line profiles: absorptive spectropolarimetric effects and temporal variability in Post-AGB, Herbig Ae/Be and other stellar types.}
\author{D. M. Harrington \& J.R. Kuhn}
\affil{Institute for Astronomy, University of Hawaii, Honolulu-HI-96822}
\email{dmh@ifa.hawaii.edu, kuhn@ifa.hawaii.edu}
\vspace{-3mm}

\begin{abstract}

We show here that the absorptive H$_\alpha$ polarized line profile previously seen in many Herbig Ae/Be (HAeBe) stars is a nearly ubiquitous feature of other types of embedded or obscured stars. This characteristic 1\% linear polarization variation across the absorptive part of the H$_\alpha$ line is seen in Post-AGB stars  as well as RV-Tau, $\delta$-Scuti, and other types. Each of these stars shows evidence of obscuration by intervening circumstellar hydrogen gas and the polarization effect is in the absorptive component, consistent with an optical pumping model. We present ESPaDOnS spectropolarimetric observations of 9 post-AGB and RV-Tau types in addition to many multi-epoch HiVIS observations of these targets. We find significant polarization changes across the H$_\alpha$ line in 8/9 stars with polarization amplitudes of 0.5\% to over 3\% (5/6 Post-AGB and 3/3 RV-Tau). In all but one of these, the polarization change is dominated by the absorptive component of the line profile. There is no evidence that subclasses of obscured stars showing stellar pulsations (RV-Tau for Post-AGB stars and $\delta$-Scuti for Herbig Ae/Be stars) show significant spectropolarimetric differences from the main class. Significant temporal variability is evident from both HiVIS and ESPaDOnS data for several stars presented here: 89 Her, AC Her, SS Lep, MWC 120, AB Aurigae and HD144668. The morphologies and temporal variability are comparable to existing large samples of Herbig Ae/Be and Be type stars. Since Post-AGB stars have circumstellar gas that is very different from Be stars, we discuss these observations in the context of their differing environments.
	
\end{abstract}


\section{Introduction}
	
	Interpreting observed linear spectropolarimetric variations across a line profile, like H$_\alpha$, has been difficult. Many phenomena such as disks, winds, magnetic fields, an asymmetric radiation field, and scattering asymmetries can produce linear polarization changes across a spectral line (cf. McLean 1979, Wood et al. 1993, Wood \& Brown 1994, Harries 2000, Ignace et al. 2004, Vink et al. 2005a, Kuhn et al. 2007). We have observed a common morphological feature in the linearly polarized H$_\alpha$ spectrum of most HAeBe stars which should be present in many other stellar systems. We confirm this expectation here and demonstrate the potential for learning about the near-star environment of embedded stars from high spectral resolution and good spectropolarimetric sensitivity ($<$0.1\% accuracy).

	There are now many detections of linear polarization signals associated with the absorptive component of the H$_\alpha$ line profile. With high spectral resolution (R$>$10000) there are comparatively large amplitude (0.2-2\%) signals in Herbig Ae/Be stars, Be's and other emission-line stars (Kuhn et al. 2007, Harrington 2008, Harrington \& Kuhn 2009). The absorptive polarization effects are ubiquitous in the Herbig Ae/Be stars with roughly 2/3 of the 'windy' and 'disky' stars showing this linear polarization signature. The classical Be stars more typically showed a broad `depolarization' morphology (10/30) but about half of these show additional absorptive effects (4/10) and another sub-set (5/30) show complex linear polarization spectral morphologies. The presence of absorptive polarization effects in most of the obscured stars in this survey suggests that the phenomenon may be present in any obscured star. In the optical pumping model described by Kuhn et al. 2007, the polarized absorption effect is present anywhere there is a difference between the line of sight to the obscuring material and the direction from the material to the star. In order to explore this hypothesis, observations of other obscured stars have been performed here. 
	
	Post-AGB stars and the associated RV-Tau type stars are often heavily obscured and have a very dynamic circumstellar environment (see van Winckel 2003 for a review). Mass-loss, shells and circumstellar envelopes are common and characteristic of the very different environments of Post-AGB, Herbig Ae/Be and Be stars. There are lower resolution spectropolarimetric studies of many different objects (cf. Beiging et al. 2006, Oppenheimer et al. 2005, Oudmaijer \& Drew 1999, Pereyra et al. 2009, Trammell 1994, Vink et al. 2002, 2003, 2005b). A comparison between spectropolarimetric morphologies and magnitudes can illuminate the important polarization mechanisms in different stellar environments.  

\begin{figure*} [!htb]
\includegraphics[width=0.23\linewidth, angle=90]{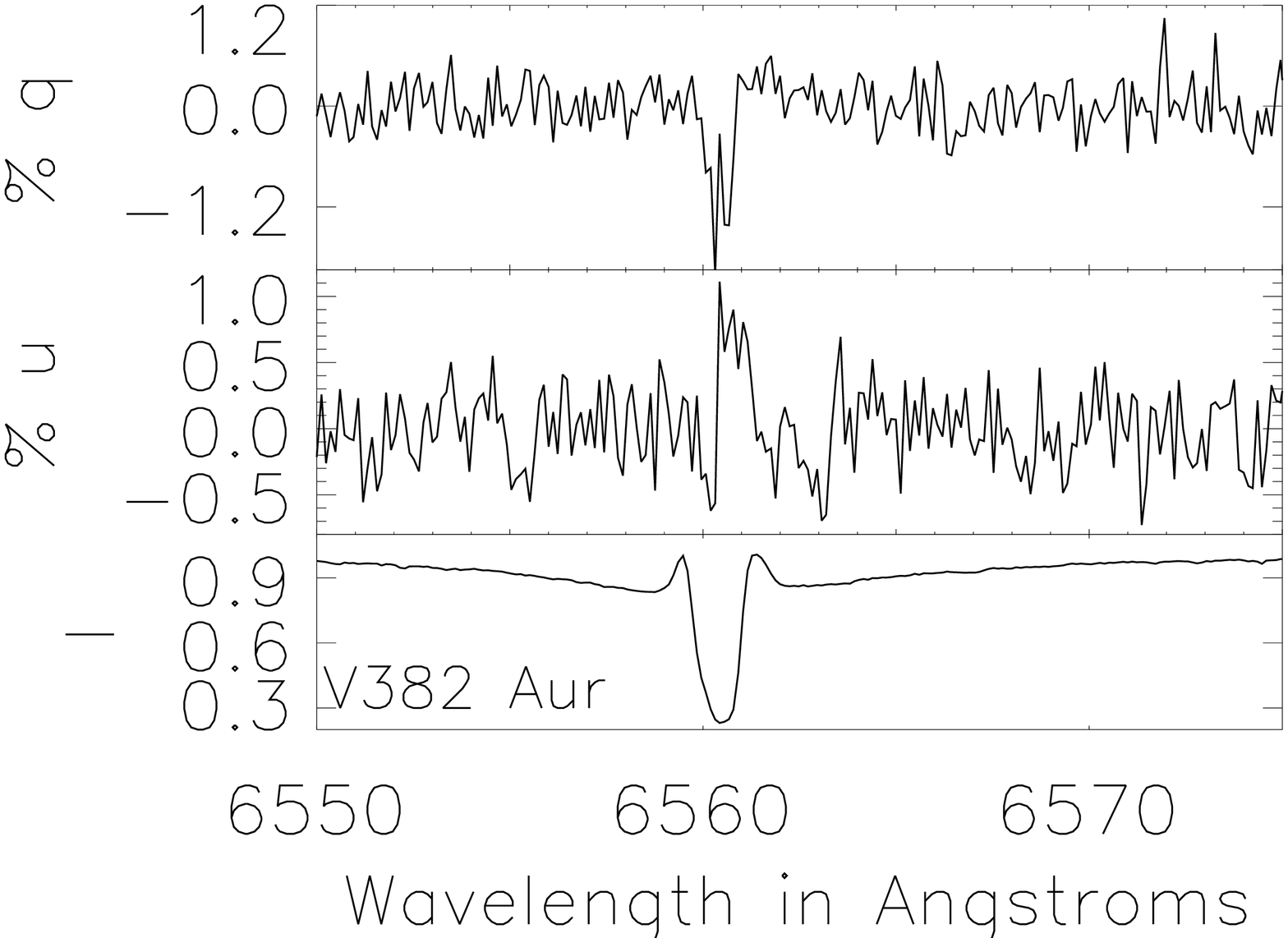}  
\includegraphics[width=0.23\linewidth, angle=90]{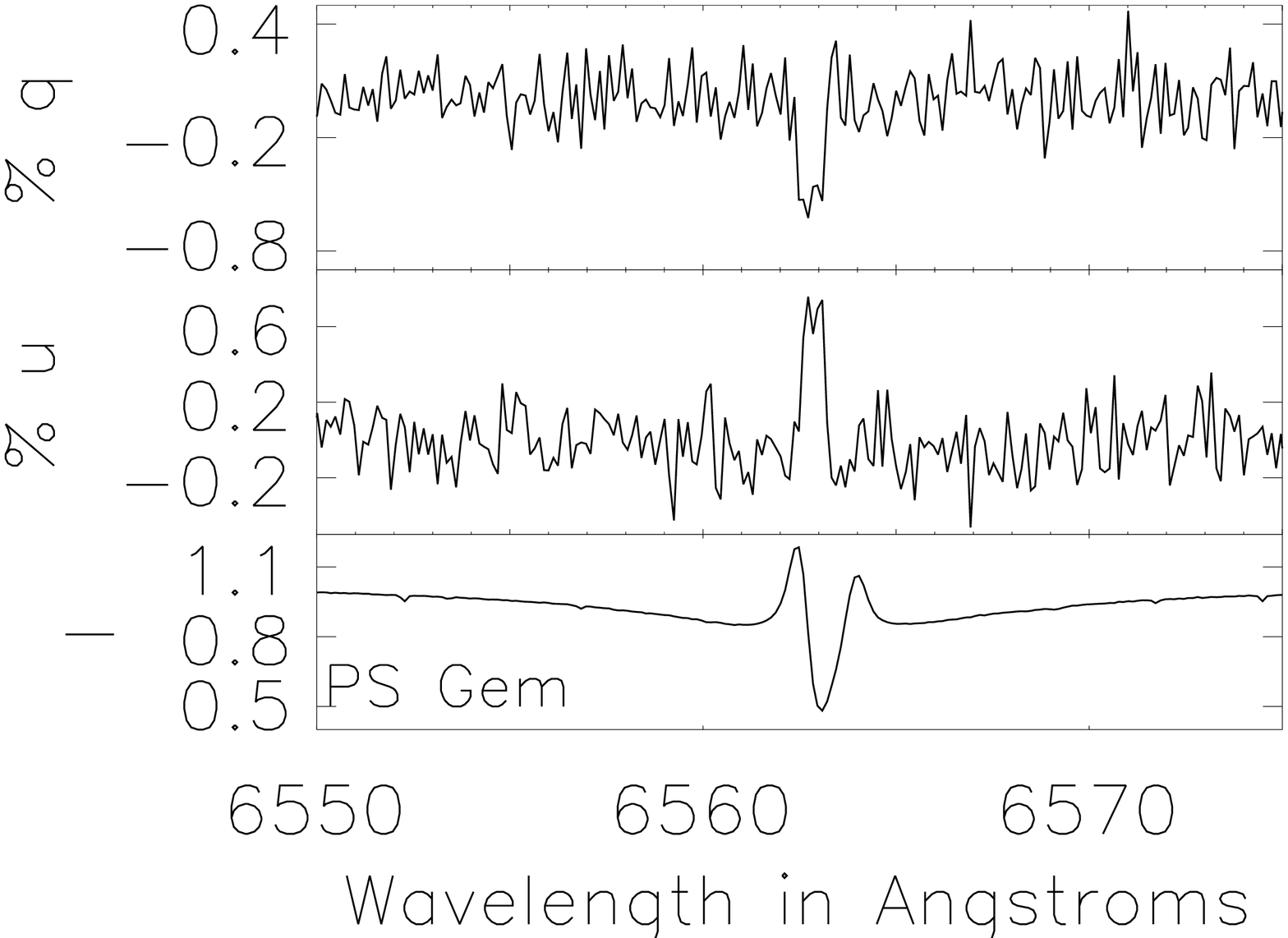} 
\includegraphics[width=0.23\linewidth, angle=90]{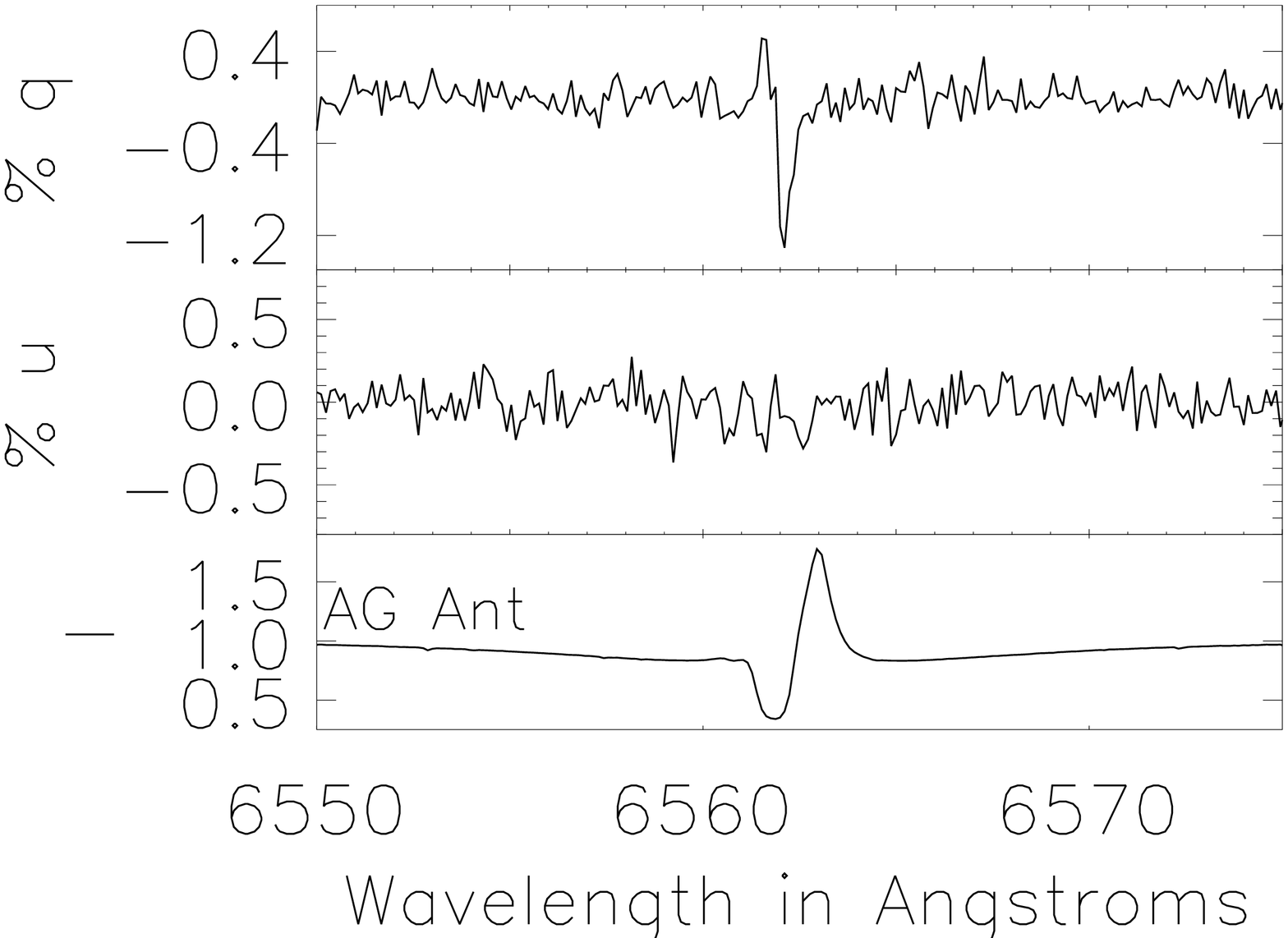} \\
\includegraphics[width=0.23\linewidth, angle=90]{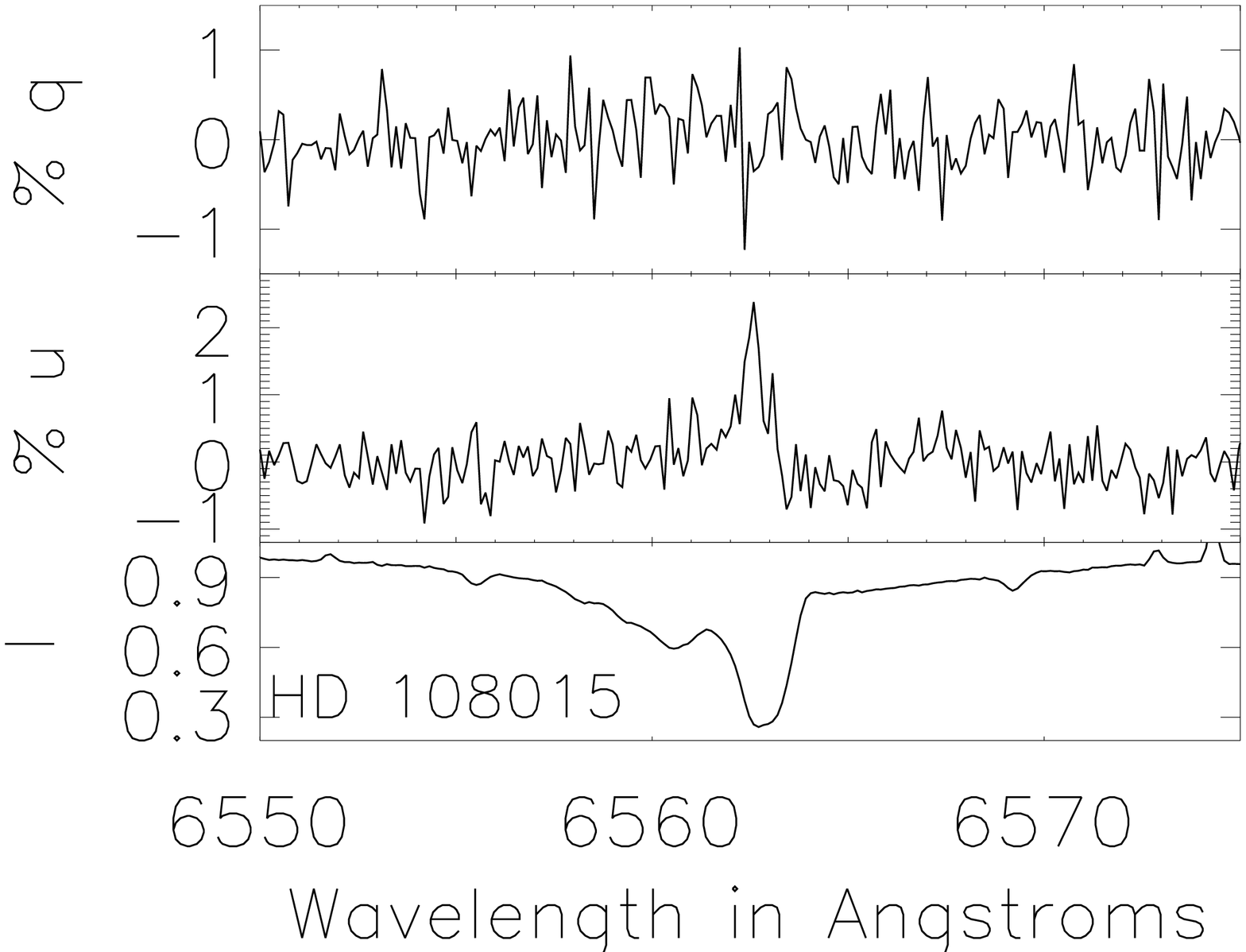}  
\includegraphics[width=0.23\linewidth, angle=90]{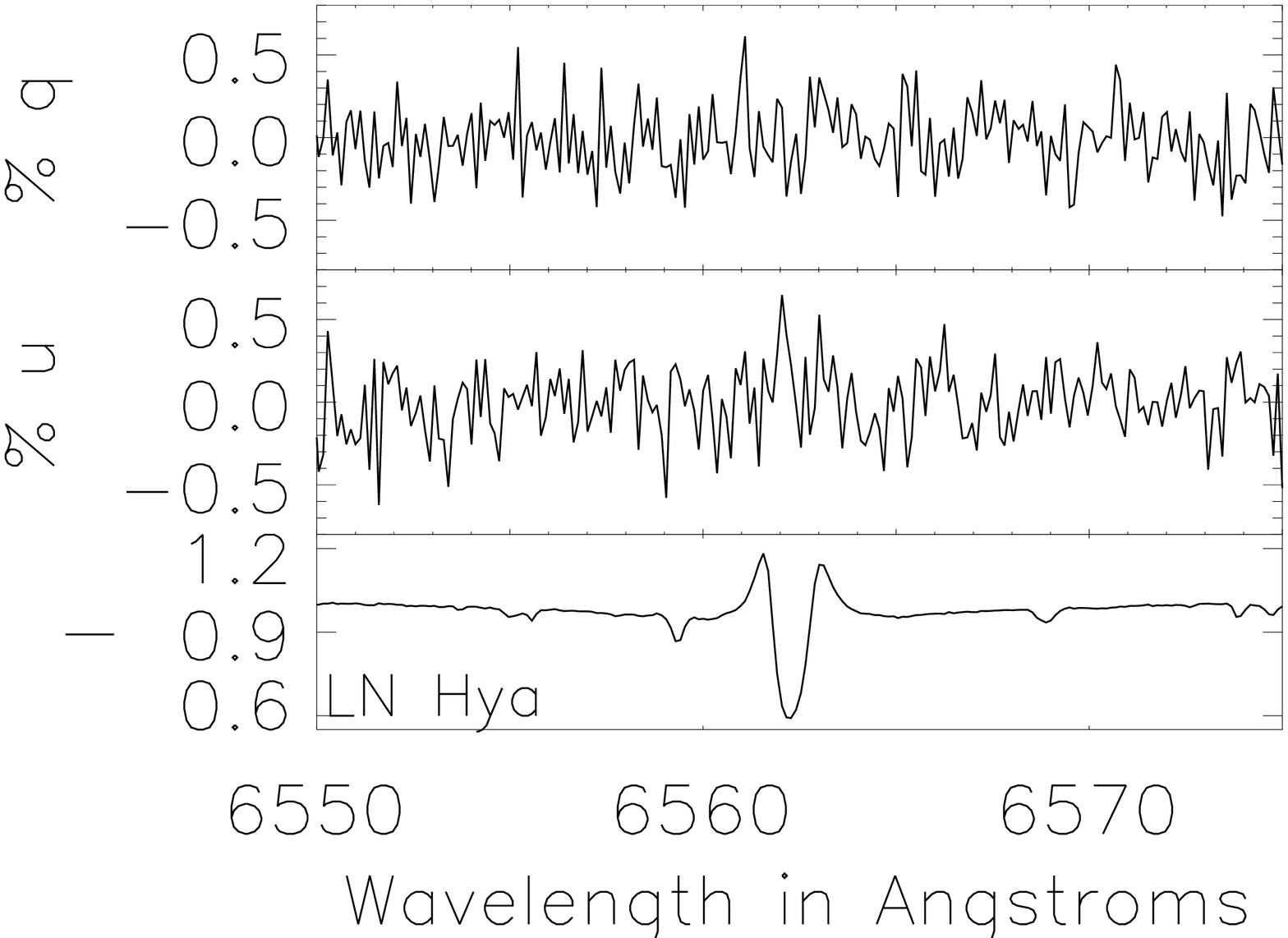}
\includegraphics[width=0.23\linewidth, angle=90]{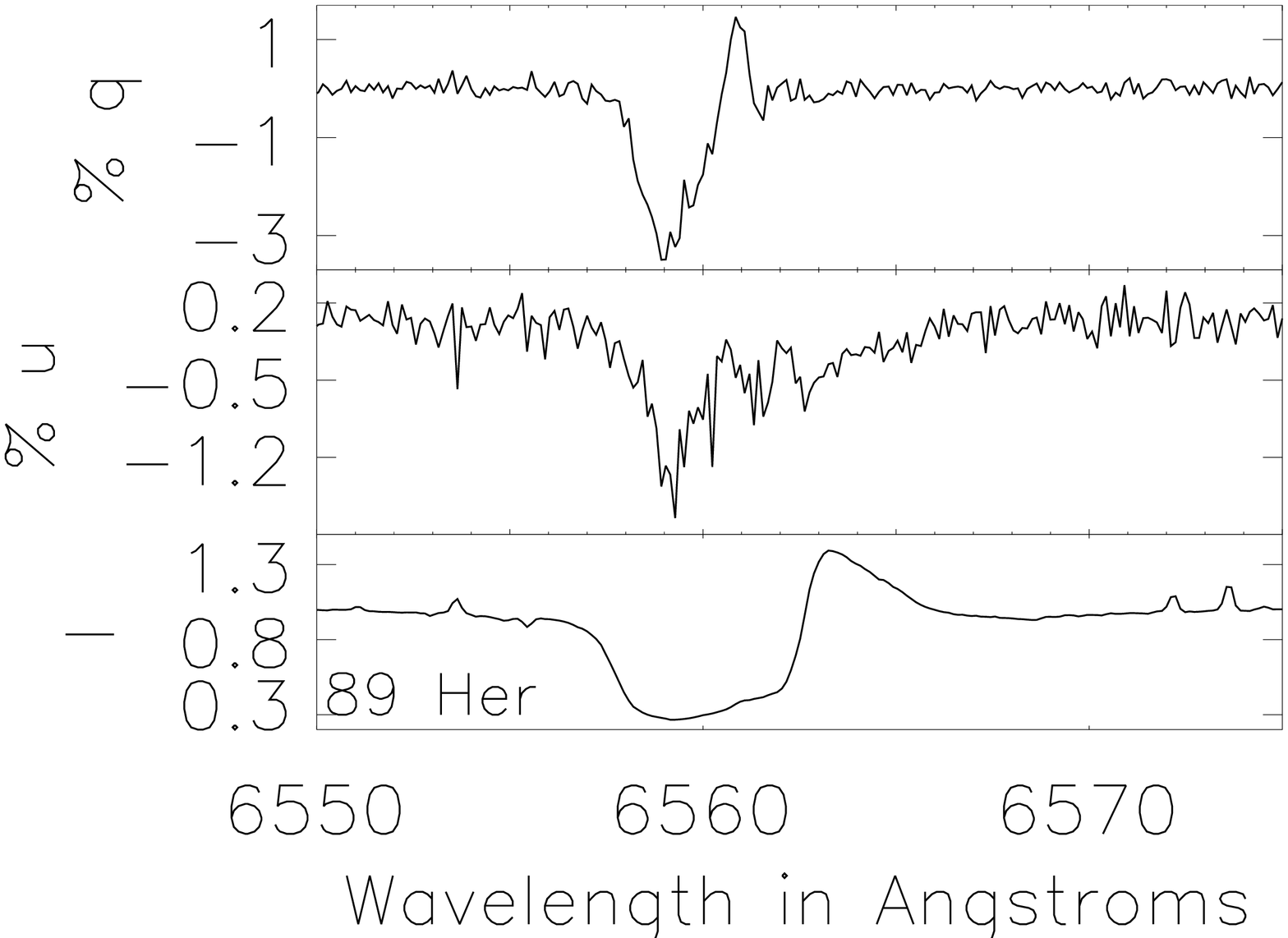} 
\caption{\label{fig:poagb} The ESPaDOnS archive H$_\alpha$ spectropolarimetry for the Post-AGB stars. From left to right - V382 Aur (HD 46703), PS Gem (HD 52961), AG Ant (HR 4049), HD 108015, LN Hya (HR 4912) and 89 Her. See table \ref{stars} for the log of observations. V832 Aur and PS Gem show a simple near-line-center excursion in a mostly-symmetric line profile. LN Hya with a similar line profile does not show a significant detection. HD 108015 has an asymmetric profile but also shows this single excursion at line center. AG Ant and 89 Her show P-Cygni type profiles but with vary different widths. 89 Her spectropolarimetry is dominated by an absorptive effects whereas AG Ant has a very narrow signature. }
\end{figure*}

\section{Observations of Post-AGB and RV-Tau Type Stars}

	Post-AGB stars are intermediate mass stars, having initial masses less than 8 or 9 M$_\odot$. Reviews of their properties can be found in Herwig 2005 and Van Winckel 2003. They are in the late stages of stellar evolution having finished their time as Asymptotic Giant Branch stars and gone through a period of very heavy mass loss. The mass loss is quite severe, being in the range of 10$^{-7}$ to 10$^{-4}$ M$_\odot$/yr with a very wide variation in the observed properties. RV-Tau type stars are a sub-type of the Post-AGB objects where the star is in the population II Cepheid instability strip at the high luminosity end (cf. Wallerstein 2002). RV-Tau type stars pulse with periods of 20 to over 50 days with alternating deep and shallow light curve minima. Most show large IR excesses and have dusty circumstellar envelopes (Gehrz 1972). Almost all RV-Tau type stars show irregularities in their light curves over time. Though the RV-Tau variables show photometric variability, the circumstellar environment is similar to the associated Post-AGB stars. This evolved stellar class was chosen for comparison with the spectropolarimetric results found in Harrington \& Kuhn 2009 for Herbig Ae/Be and Be stars because of their markedly different evolutionary status and circumstellar environment. 
	
	The observations used for this project were performed with the 3.6m Canada France Hawaii Telescope (CFHT) using the ESPaDOnS spectropolarimeter with a nominal average spectral resolution of $R=\frac{\lambda}{\delta\lambda}=68000$ and with the HiVIS spectropolarimeter on the 3.7m AEOS telescope at a spectral resolution of $R=13000$ (cf. Donati et al. 1997, 1999, Harrington et al. 2006, Harrington \& Kuhn 2008). 
	
	The Canadian Astronomy Data Centre (CADC) provides archival access to all released ESPaDOnS data. Archival ESPaDOnS observations of the stars shown in table \ref{stars} taken in 2006 were downloaded and reduced with the dedicated ESPaDOnS reduction script, Libre-ESPRIT (Donati et al. 1997). The reduction scripts output a wavelength intensity, polarization and check-spectrum for each spectral order. In order to optimize signal-to-noise and wavelength coverage, these output spectra have been binned in wavelength to regularize the spectral resolution and to combine overlapping orders. The bin size was calculated as four times median resolution element size, giving 0.12{\AA} per wavelength bin or a single-pixel-sampling resolution of 54600 at H$_\alpha$. The intensity, spectropolarimetry and wavelength data in each spectral bin were averaged resulting in a monotonically increasing, nearly-regular wavelength coverage. In the H$_\alpha$ region for this detector, two orders completely cover the line. The unbinned spectral-pixel sizes are 0.036 and 0.044{\AA} for orders 34 and 35 (6470 and 6660{\AA}). The regularization procedure typically averages 5-6 spectral-pixels giving a 2-fold improvement in signal-to-noise compared to the Libre-Esprit output shown in table 1. All observations had continuum signal-to-noise ratio's from the Libre-Esprit scripts of 200-1100. After the binning, a polarimetric accuracy of 0.05\% to 0.2\% is achieved. 

\begin{table}[!h,!t,!b]
\begin{center}
\caption{Stellar Properties \& ESPaDOnS Observations \label{stars}}
\vspace{2mm}
\begin{tabular}{lccclcrr}
\hline
\hline
{\bf Name}        	& {\bf HD}   	& {\bf V}	& {\bf Sp}		&{\bf Type} 	& {\bf UT} & {\bf S/N} 	& {\bf \% P} \\ 
\hline
\hline
V382 Aur          	&46703           	& 9.1     	& F7 IVw      	& PAGB       	& 09  	& 284       	& 1.5\% \\  
PS Gem            	&52961         	& 7.4      	& A0              	& PAGB      	& 07  	& 554       	& 0.8\%  \\
AG Ant               	&89353           	& 5.5     	& B9.5 Ib-II   	& PAGB      	& 07  	& 741       	& 1.2\%  \\
HD 108015       	&108015        	& 8.0     	& F4 Ib/II       	& PAGB      	& 08   	& 245        & 1.5\%  \\
LN Hya             	&112374        	& 6.7     	& F3 Ia          	& PAGB     	& 08   	& 387        &  ---        \\
89 Her               	&163506        	& 5.5      	& F2 Ibe      	& PAGB       	& 07   	& 596        &3.4\%   \\ 
\hline
U Mon               	&59693         	& 6.8     	& K0 Ibpv       	& RVTau     	& 07   	& 1108   	&0.5\%  \\ 
U Mon                	&59693         	& 6.8     	& K0 Ibpv       	& RVTau     	& 08    	& 1140    	&0.5\%  \\ 
RU Cen            	&105578       	& 9.0     	& G2 w          	& RVTau      	& 09    	& 291     	& 0.7\%   \\ 
AC Her              	&170756        	& 7.6     	& F4 Ibpv       	& RVTau     	& 07    	& 607    	&0.8\%   \\  
\hline
\hline
\end{tabular}
\end{center}
The table of stellar properties and ESPaDOnS observations. The columns list the stars common name, HD catalog number, V magnitude, spectral type (Sp) and star type as listed in the Simbad database. For each star, the February 2006 UT date of observation (UT), signal-to-noise (S/N) and the peak degree of polarization (\%P) measured across the H$_\alpha$ line is shown. The signal-to-noise is calculated as the average S/N value output by the Libre-Esprit package for the ccd-pixel I spectrum for both q and u using both orders 34 and 35 (6470{\AA} and 6660{\AA}). For the H$_\alpha$ line there is complete wavelength coverage and order overlap across the entire line in both orders. The binning procedure we apply results in at least a doubling of the signal-to-noise for each measurement listed in the S/N column.
\end{table}

\begin{table}[!h,!t,!b]
\begin{center}
\caption{HiVIS Post-AGB \& RV-Tau Observations \label{hivisobs}}
\vspace{2mm}
\begin{tabular}{lcccrr}
\hline
\hline
{\bf Name}& {\bf Date}	& {\bf Time}	& {\bf ET}	&{\bf Az} 	& {\bf El} 	 \\ 
		& {\bf UT}      	& {\bf UT}  	& {\bf s}   	&	     	&  		 \\ 
\hline
\hline
89 Her	& 	6-25		& 12 02	& 240		& 286	& 62	\\
89 Her	&	6-27		& 06 58	& 420		& 73		& 48	\\
89 Her	&	7-24		& 12 02	& 420		& 287	& 37	\\
89 Her	&	7-25		& 12 38	& 600		& 289	& 28	\\
89 Her	&	7-26		& 10 16	& 600		& 287	& 61	\\
89 Her	&	7-27		& 10 35	& 600		& 286	& 56	\\
\hline
PS Gem	&	9-28		& 13 30	& 480		& 92		& 38	\\
PS Gem	&	9-28		& 14 12	& 480		& 97		& 50	\\
\hline
U Mon	&	9-28		& 14 42	& 600		& 123	& 39	\\
\hline 
AC Her	&	6-27		& 11 50	& 600		& 277	& 71	\\
AC Her	&	6-27		& 12 32	& 600		& 277	& 62	\\
AC Her	&	7-24		& 12 48	& 420		& 283	& 33	\\
AC Her	&	7-26		& 11 00	& 600		& 270	& 59	\\
AC Her	&	7-26		& 12 25	& 600		& 282	& 40	\\
AC Her	&	7-27		& 11 07	& 600		& 279	& 54	\\
\hline
\hline
SS Lep	&	9-27		& 12 50	& 480		& 124	& 29	\\
SS Lep	&	9-27		& 13 24	& 480		& 130	& 36	\\
SS Lep	&	9-28		& 12 22	& 480		& 120	& 24	\\
SS Lep	&	9-28		& 12 56	& 480		& 126	& 32	\\	
\hline
\hline
\end{tabular}
\end{center}
The table of HiVIS observations. The columns list the stars common name, the UT date and time, the exposure time in seconds (ET) and the pointing of the AEOS telescope in azimuth and elevation (Az, El). The pointing and exposure-time information is included because the polarimetric properties of the telescope are pointing-dependent and the polarimetric reference frame rotates with time. See Harrington et al. 2006 and Harrington \& Kuhn 2008 for details.
\end{table}

	The HiVIS observations were taken between June and September 2008 and were processed with the HiVIS dedicated package. See Harrington \& Kuhn 2008 for the data reduction process and detailed instrument information. The observations were reduced using the simple-average method (not the optimal-spectral-extraction) and were not binned-by-flux. The observations were linearly binned by a factor of five to improve the signal-to-noise. This gives us a regular wavelength array to match the ESPaDOnS observations. It should be pointed out that in the R=13000 mode, thorium-argon lamp calibrations show a full-width-half-max of 16 spectral pixels. Even binning by a factor of five gives full spectral resolution with a sampling of 3 spectral pixels per resolution element. The Pixis CCD, described in Harrington \& Kuhn 2008, has been replaced with an Apogee Alta 3k$^2$ array for these observations giving much larger wavelength coverage, though at the expense of reduced quantum efficiency. The log of HiVIS observations and telescope pointings is shown in table \ref{hivisobs}.

	The AEOS telescope does unpredictable (so far) damage to the angular phase of the polarization we measure. While the telescope-induced Q and U phase angle appears to be only a function of wavelength, telescope altitude, and azimuth, we have not yet settled on a deterministic correction for all pointings. Thus some of our HiVIS data in the discussions below is rotated empirically to allow temporal variability comparison. A simple rotation between HiVIS and ESPaDOnS observations was shown to make HiVIS reproduce ESPaDOnS spectropolarimetric morphologies, though with the possibility of a reduction in magnitude (Harrington \& Kuhn 2008, Harrington 2008). We also note that this effect limits our exposure times for near-zenith telescope pointings since it can introduce unacceptable linear polarization variability due to the rapidly changing mirror orientations near zenith (in this alt-az telescope).

	There are observations of six Post-AGB stars and three RV-Tau type variable stars available in the ESPaDOnS archive. The names, spectral types and magnitudes from Simbad and the observational details are shown in table \ref{stars}. The stars range from spectral type K0 to B9.5. The H$_\alpha$ spectra of these objects are quite varied. The ESPaDOnS spectropolarimetry for the Post-AGB stars is shown in figure \ref{fig:poagb}. 

\begin{figure} [!h]
\begin{center}
\includegraphics[width=0.55\linewidth, angle=90]{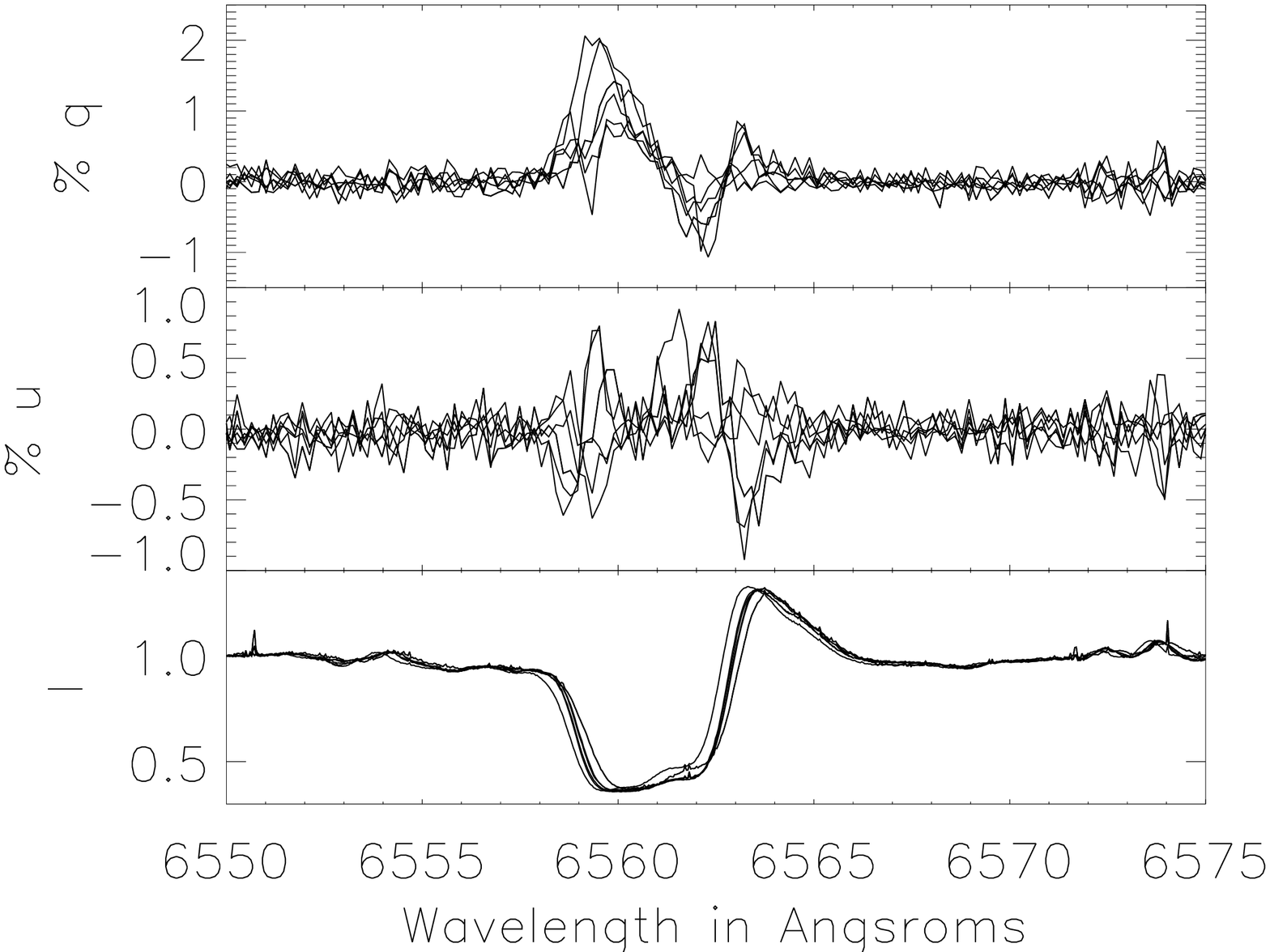}  
\includegraphics[width=0.55\linewidth, angle=90]{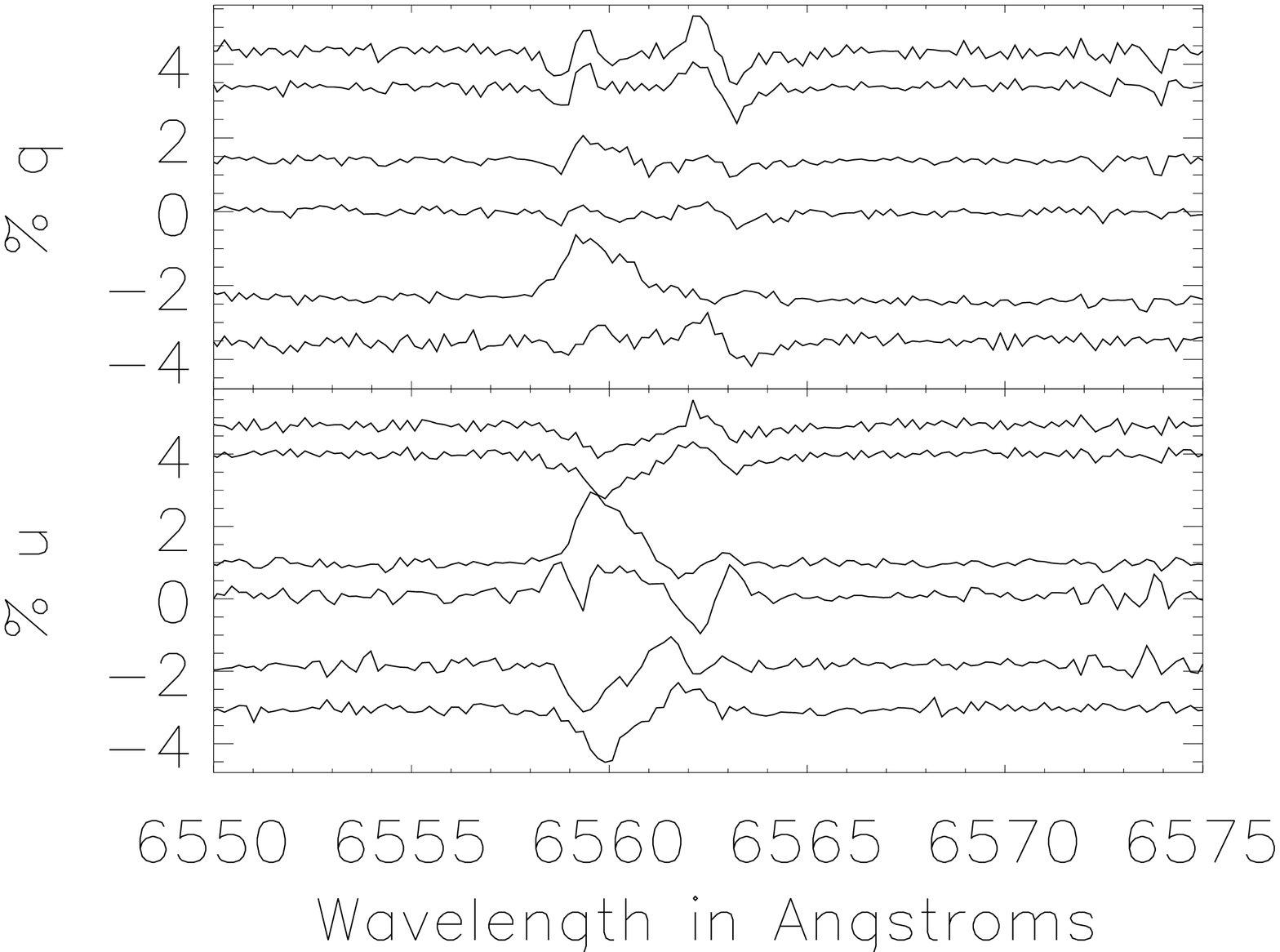}  
\caption{\label{fig:hivis89her} The HiVIS H$_\alpha$ spectropolarimetry for the Post-AGB star 89 Her for autumn 2008. The top box shows 89 Her where the polarization reference frame of each spectrum has been rotated to maximize q at 6560{\AA}  above the continuum-normalized intensity spectrum. There is clearly significant morphological variability. The bottom box shows the q and u spectra in the HiVIS polarization reference frame. There are clear broad blue-shifted detections in every epoch of observations. Time increases vertically. The observing log is shown in table \ref{hivisobs}. There is a noticeable similarity in the observations of June 25th and July 26th and 27th. All three observations were taken near the same pointing.}
\end{center}
\end{figure}
	
	Two stars show clear P-Cygni type profiles (AG Ant and 89 Her). 89 Her is complex in morphology and very large in amplitude. AG Ant has a relatively simple spectropolarimetric signature, being almost a simple narrow antisymmetric signature. The P-Cygni type profile is similar to that observed by others (cf. Bakker et al. 1998). Three stars show central and mostly symmetric emissive and absorptive components - PS Gem, V382 Aur and LN Hya. Two of these, V382 Aur and PS Gem, show relatively simple monotonic excursions in q and u and the third is a non-detection. In all cases there is emission with overlying absorptive effects and the polarization changes only in the absorptive components. The two stars with strongly blue-shifted absorption range in spectral type from F2 to B9 while the three stars with more central components span F4 - A0. The final star with a unique H$_\alpha$ profile, HD 108015, has an asymmetric intensity profile but a simple central deviation in Stokes u. All the detections except 89 Her span a narrow wavelength range.

\begin{figure} [!h]
\begin{center}
\includegraphics[width=0.55\linewidth, angle=90]{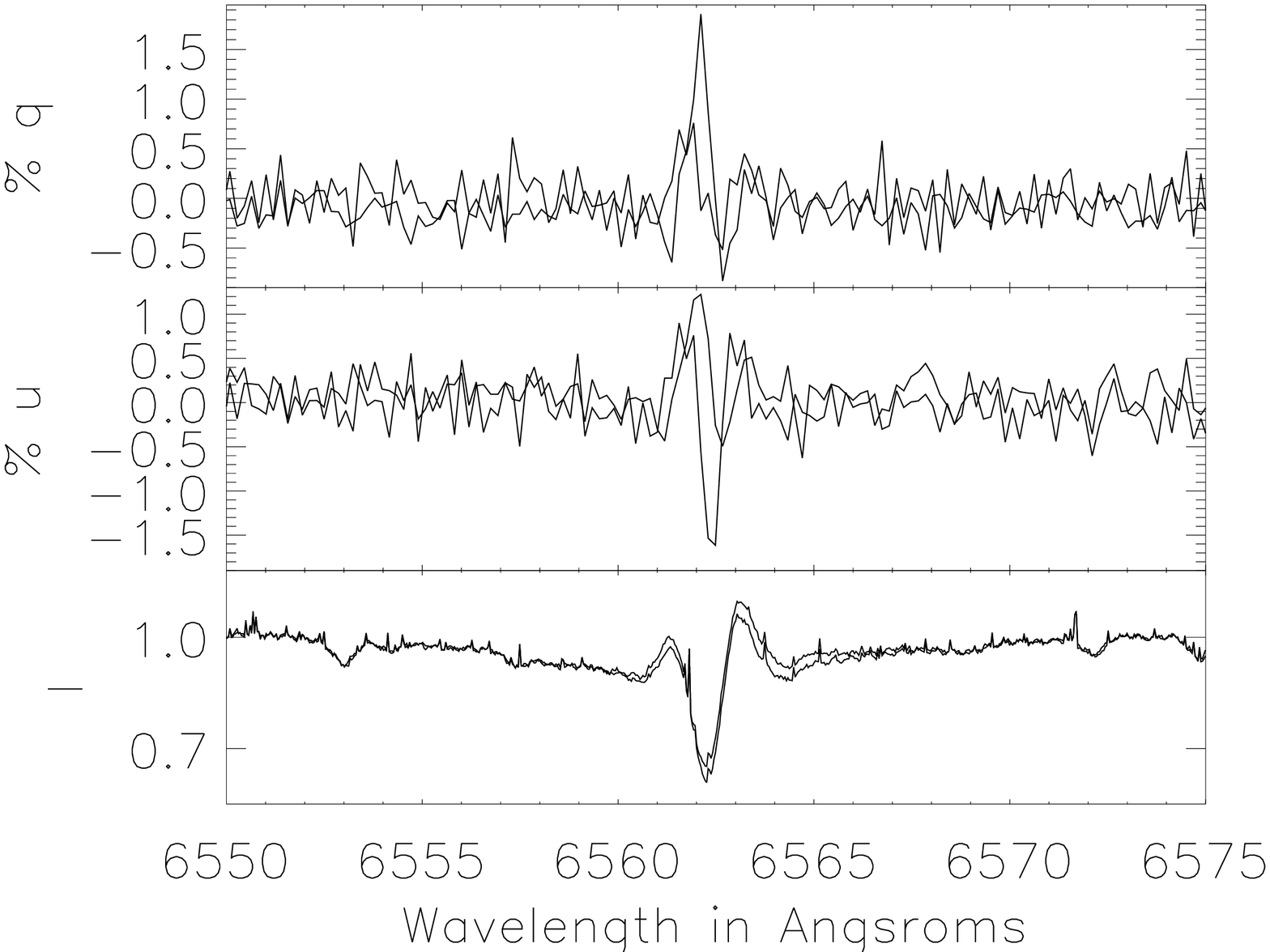}  
\includegraphics[width=0.55\linewidth, angle=90]{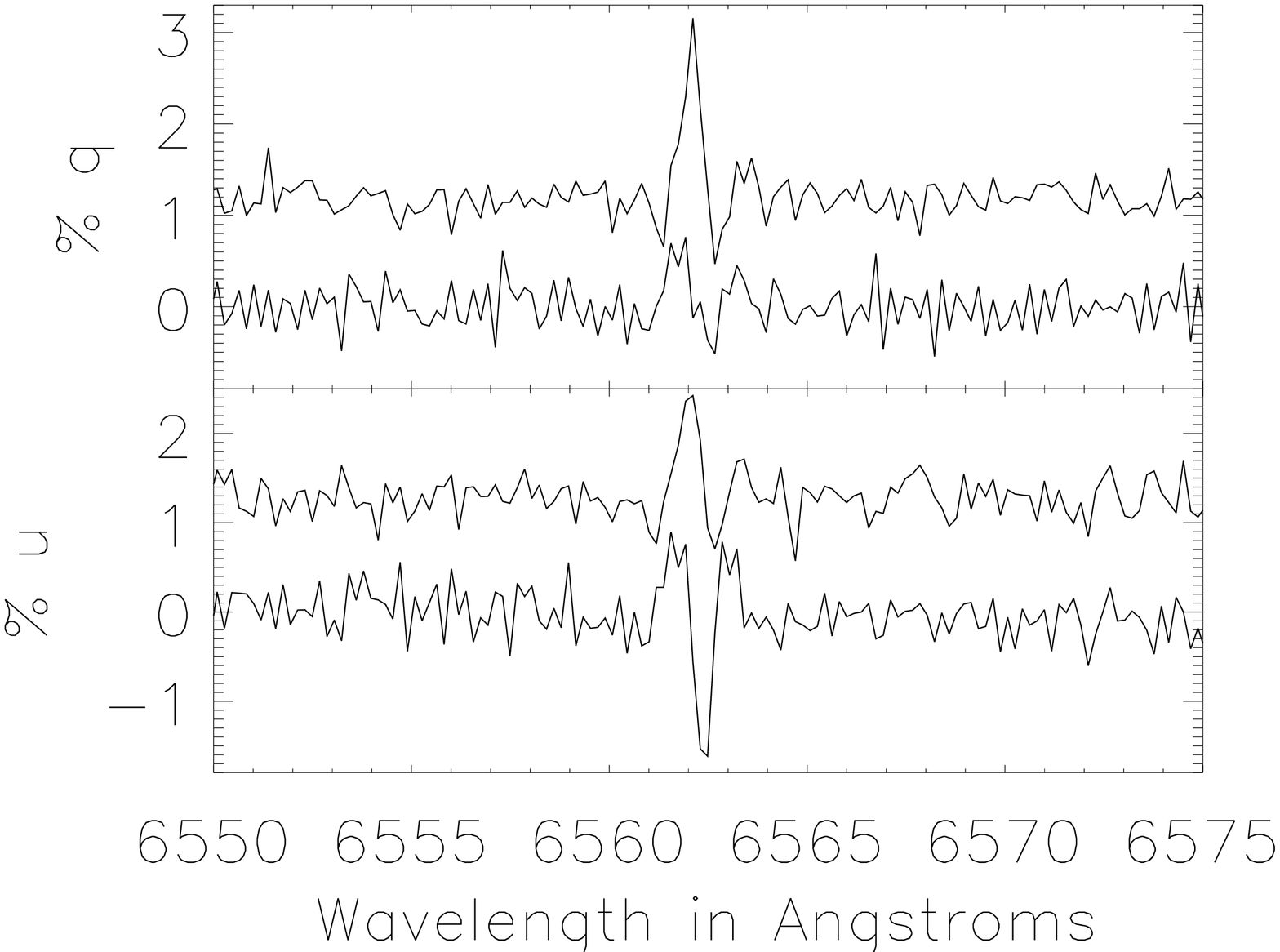}  
\caption{\label{fig:hivispsgem} The HiVIS H$_\alpha$ spectropolarimetry for the Post-AGB star PS Gem for autumn 2008. The top panel shows the unrotated q and u spectra with corresponding continuum-normalized intensity spectrum. The bottom box shows the unrotated q and u spectra. They show significant differences, however this could be to the rotational differences between observations. Time increases vertically. The observing log is shown in table \ref{hivisobs}.}
\end{center}
\end{figure}
	
	HiVIS observations of 89 Her and PS Gem are shown in figures \ref{fig:hivis89her} and \ref{fig:hivispsgem}. There are clear large-amplitude detections in every observation of 89 Her. Since the HiVIS polarization properties change with pointing as well as the polarimetric reference frame, an arbitrary least-squares rotation is applied to some data sets as indicated throughout this paper in order to align the polarized spectra. See Harrington \& Kuhn 2008 for details. The rotation angle was determined for 89 Her by maximizing the q spectrum 6559-6561\AA. 
	
	After the rotational alignment, as seen in the top panel of figure \ref{fig:hivis89her}, there is still very significant morphological variation in the u spectrum. The AEOS telescope polarization properties can rotate a spectropolarimetric signature or reduce it's detected magnitude, but they cannot induce an effect. With the rotational alignment applied to the data set, the significant variability in the morphology of the u spectrum can be clearly seen spanning almost the entire line though with a clear blue-shift. The bottom panel of figure \ref{fig:hivis89her} shows the observations in the unrotated instrument frame. The signatures are seen to be very significant and complex in q and u. In that plot, time increases vertically. There are two observations from June as the bottom two spectra showing similar u but significantly different q. However, these two were taken at quite different pointings. The July observations were taken at essentially the same azimuth but with elevations from 30-60$^\circ$. The q spectra are quite similar but with a large variation in u. There is a noticeable rough similarity between the June 25th observations (bottom) and the July 26th and 27th observations (top two). These observations were all taken at essentially the same pointing. Though there are significant differences when examined in detail, these observations show that the overall form of the signature is stable over one month time-scales even though there is very significant variability between all observations.

\begin{figure} [!h]
\begin{center}
\includegraphics[width=0.55\linewidth, angle=90]{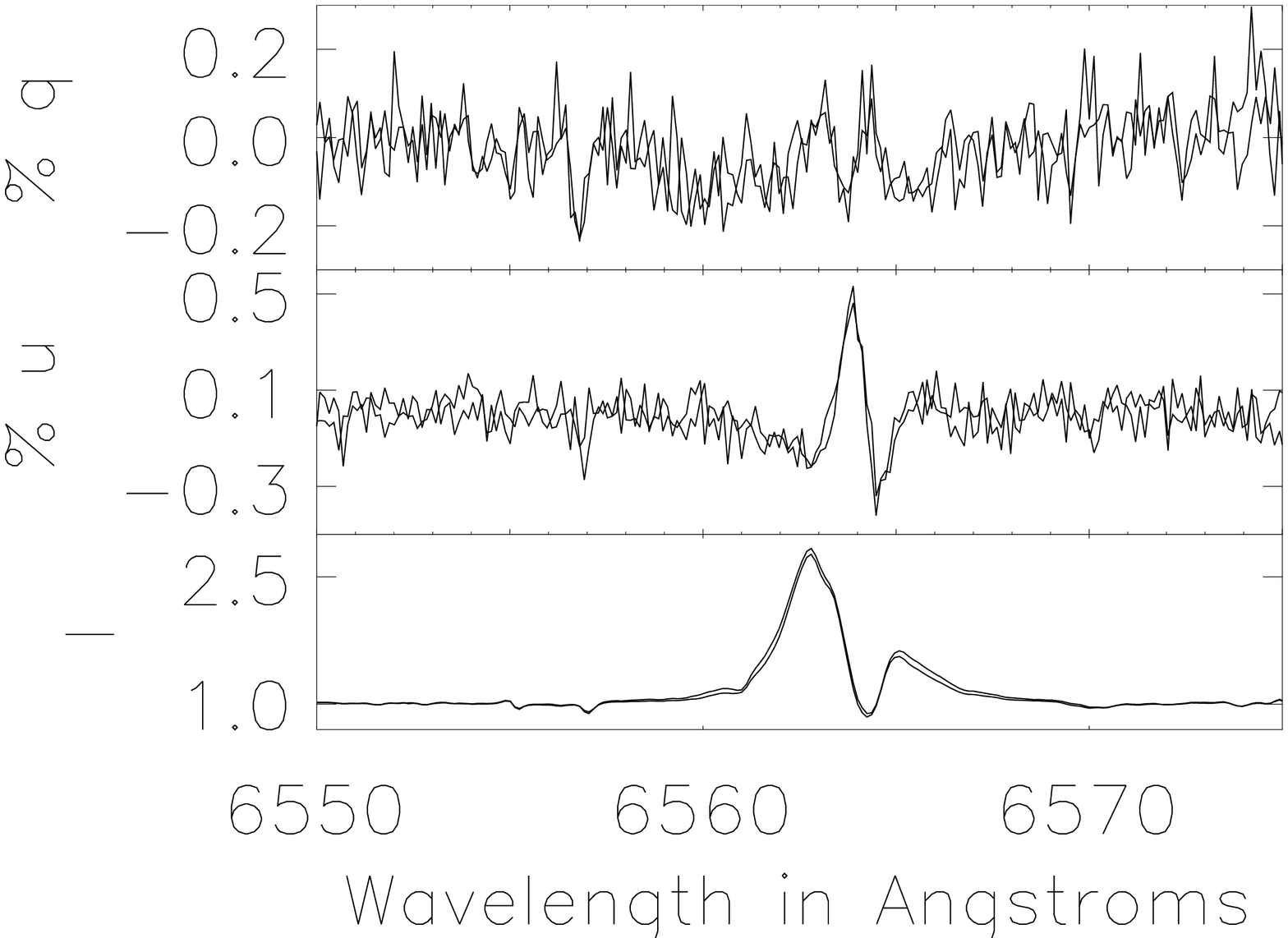}  
\includegraphics[width=0.55\linewidth, angle=90]{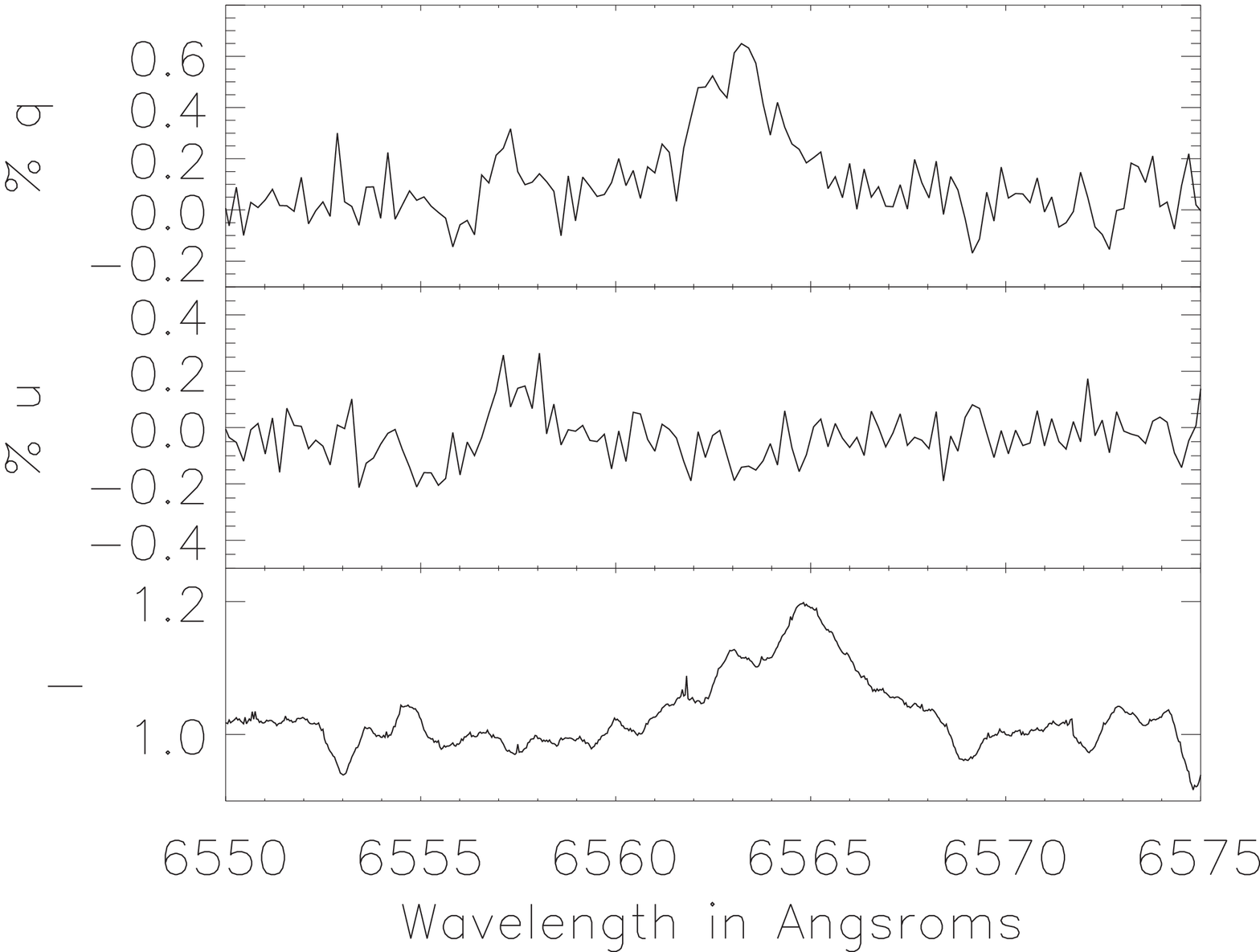} 
\caption{\label{rvtau-umon} The H$_\alpha$ spectropolarimetry for the RV-Tau type star U Mon. The top plot shows the ESPaDOnS archive spectropolarimetry from February 7th and 8th 2006. The bottom plot shows HiVIS spectropolarimetry from September 28th 2008. The H$_\alpha$ line profile changed dramatically between the two observations.}
\end{center}
\end{figure}

	The HiVIS observations of PS Gem are shown in figure \ref{fig:hivispsgem}. They were taken roughly one hour apart with a change in azimuth of 5$^\circ$ and an elevation change of 12$^\circ$. A small but clear difference is seen in the H$_\alpha$ intensity profile. No rotation has been applied to this data set as there is no broad wavelength range in the two observations over which to define a clear maximum for rotational alignment. The top panel of the figure shows the q u and intensity spectra in one graph. The q and u spectra are shown vertically separated in the bottom panel to illustrate the temporal changes. There is a clear narrow detection in both data sets with a noticeably more complex morphology than the ESPaDOnS observations of figure \ref{fig:poagb}. There is an obvious antisymmetry to the top q and bottom u spectra. The apparent variability could possibly be due to changing rotation frame and telescope induced rotation and more observations of this target are necessary to conclude that this star is variable on hour time-scales.

\begin{figure} [!h]
\begin{center}
\includegraphics[width=0.55\linewidth, angle=90]{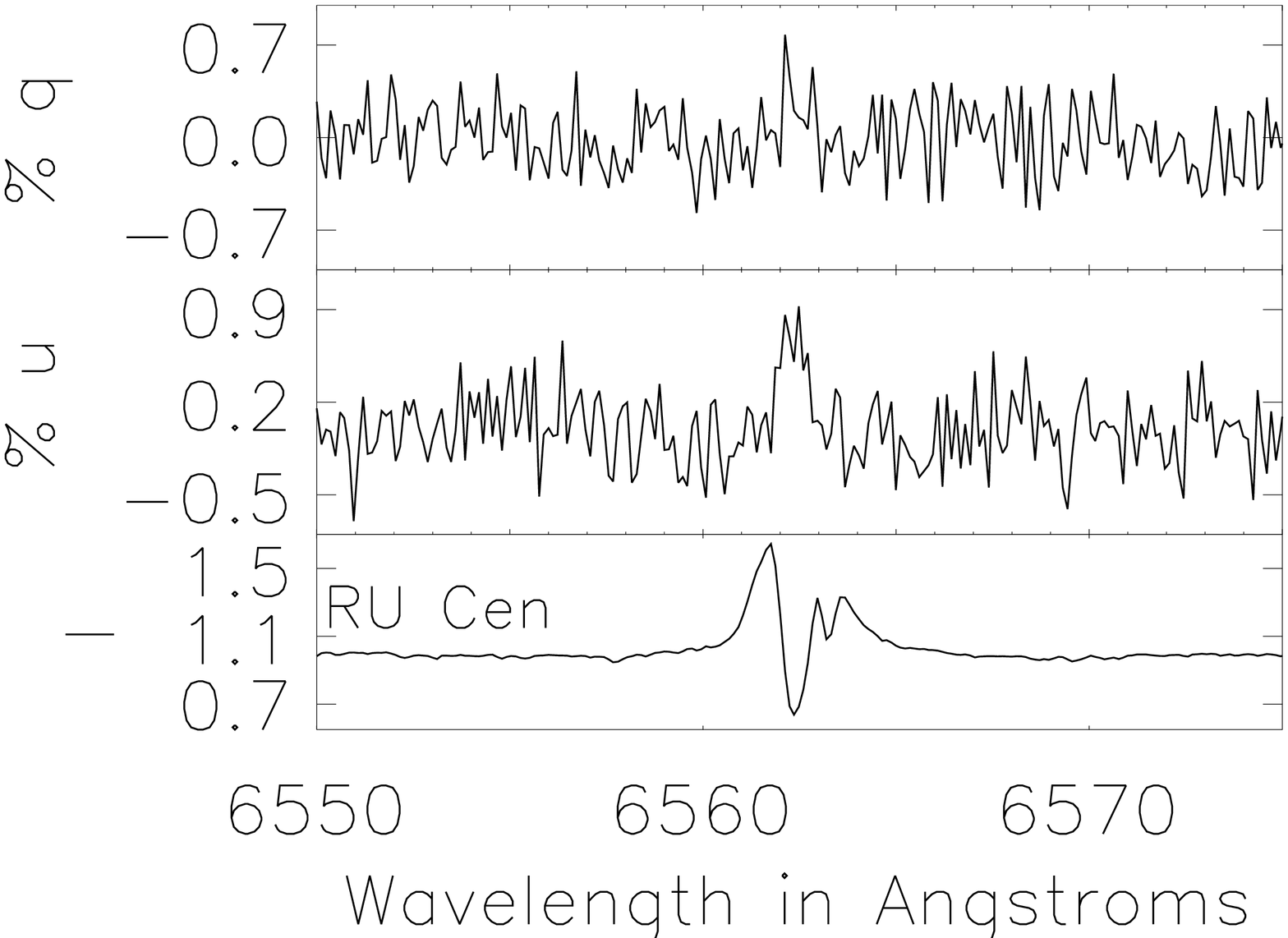}  
\includegraphics[width=0.55\linewidth, angle=90]{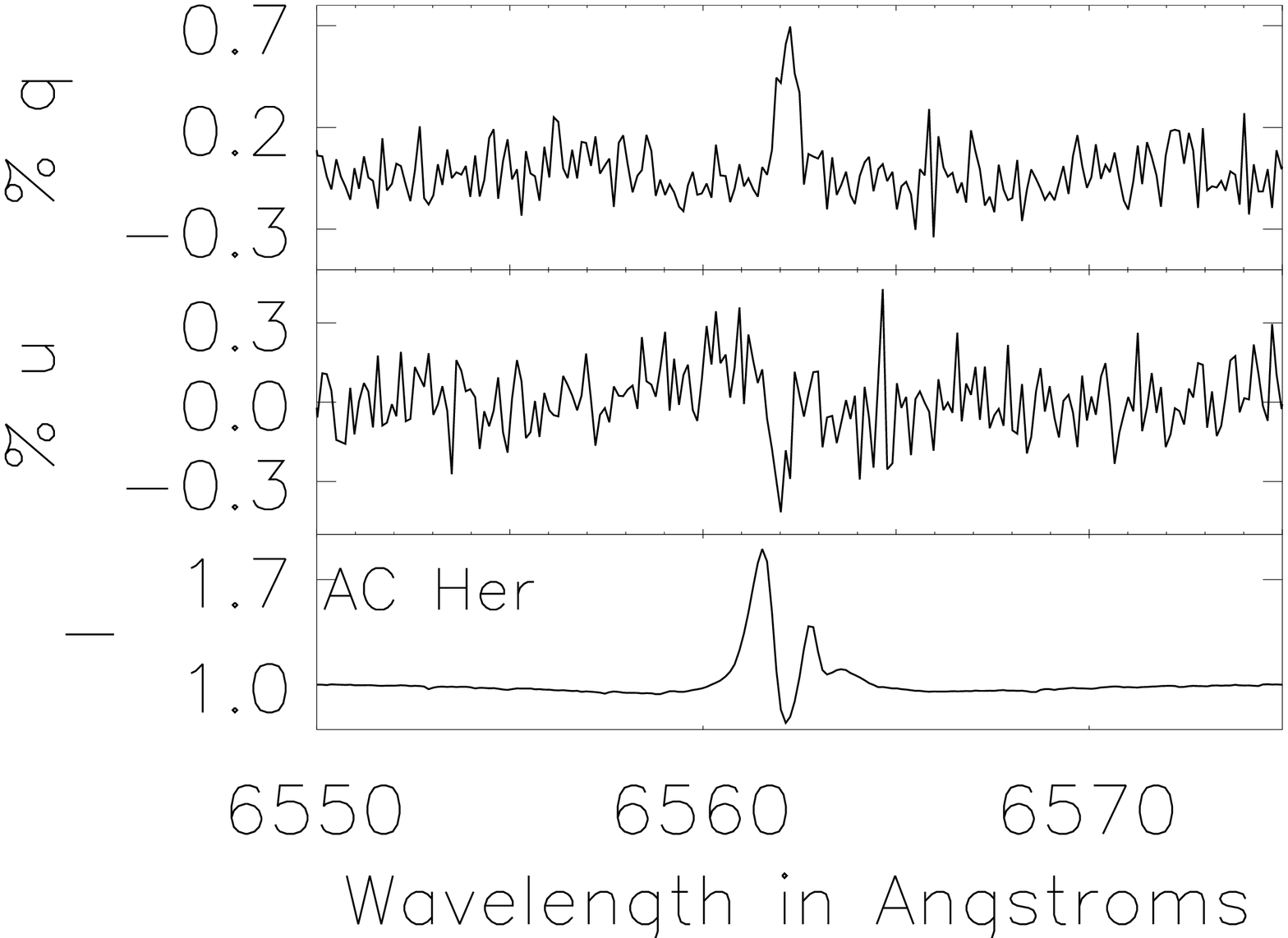}
\caption{\label{rvtau-esp} The ESPaDOnS archive H$_\alpha$ spectropolarimetry for the RV-Tau type stars RU Cen (top) and AC Her (bottom). Observations were taken in February 2009. The H$_\alpha$ profiles are similar at this epoch, showing moderate emission with a fairly complex overlying absorption and a net red-shift to that absorption. The qu-morphology is essentially simple excursions in qu-space. However, the morphology was not constant in time for AC Her - see text for details.}
\end{center}
\end{figure}

\begin{figure} [!h]
\begin{center}
\includegraphics[width=0.55\linewidth, angle=90]{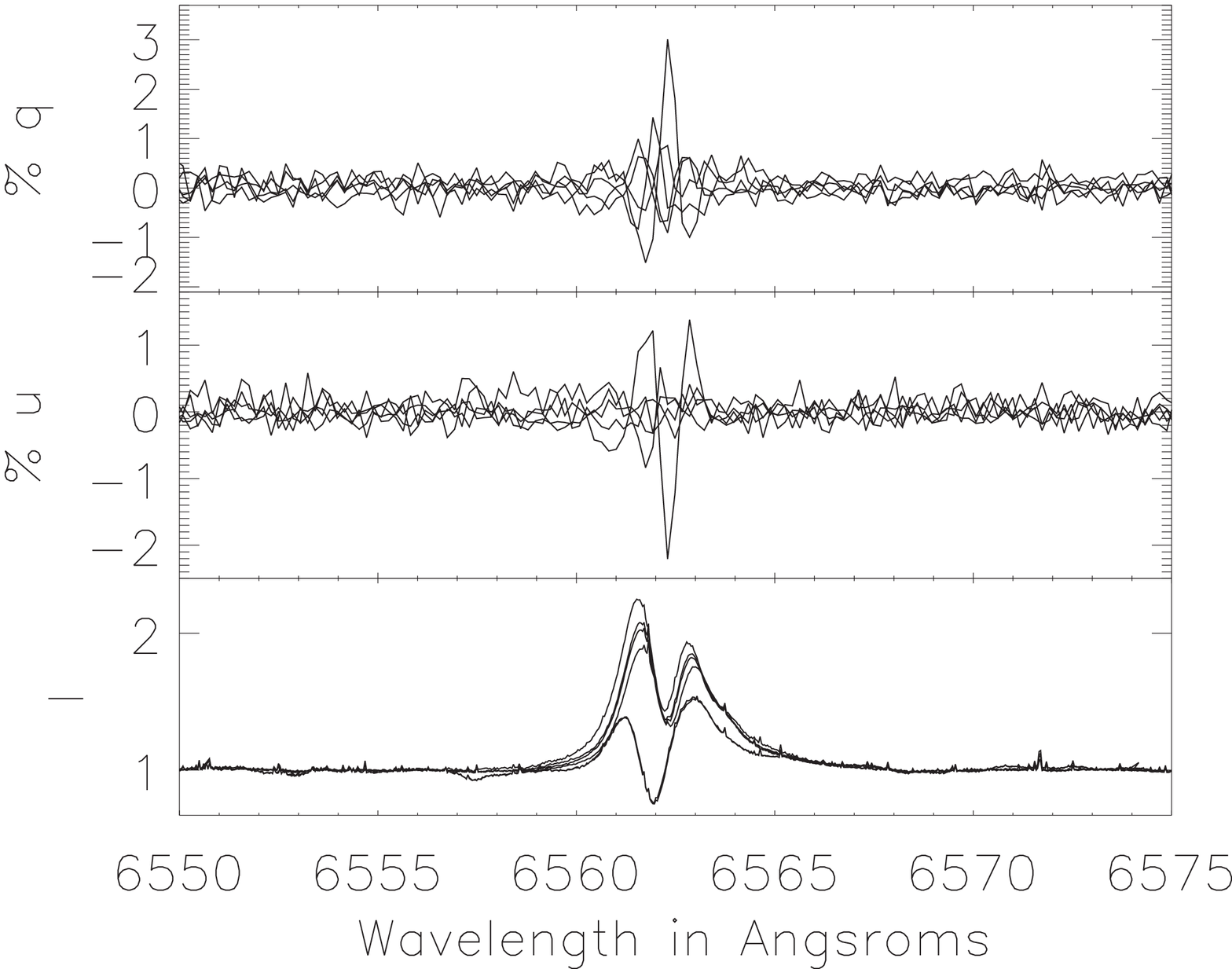}   
\includegraphics[width=0.55\linewidth, angle=90]{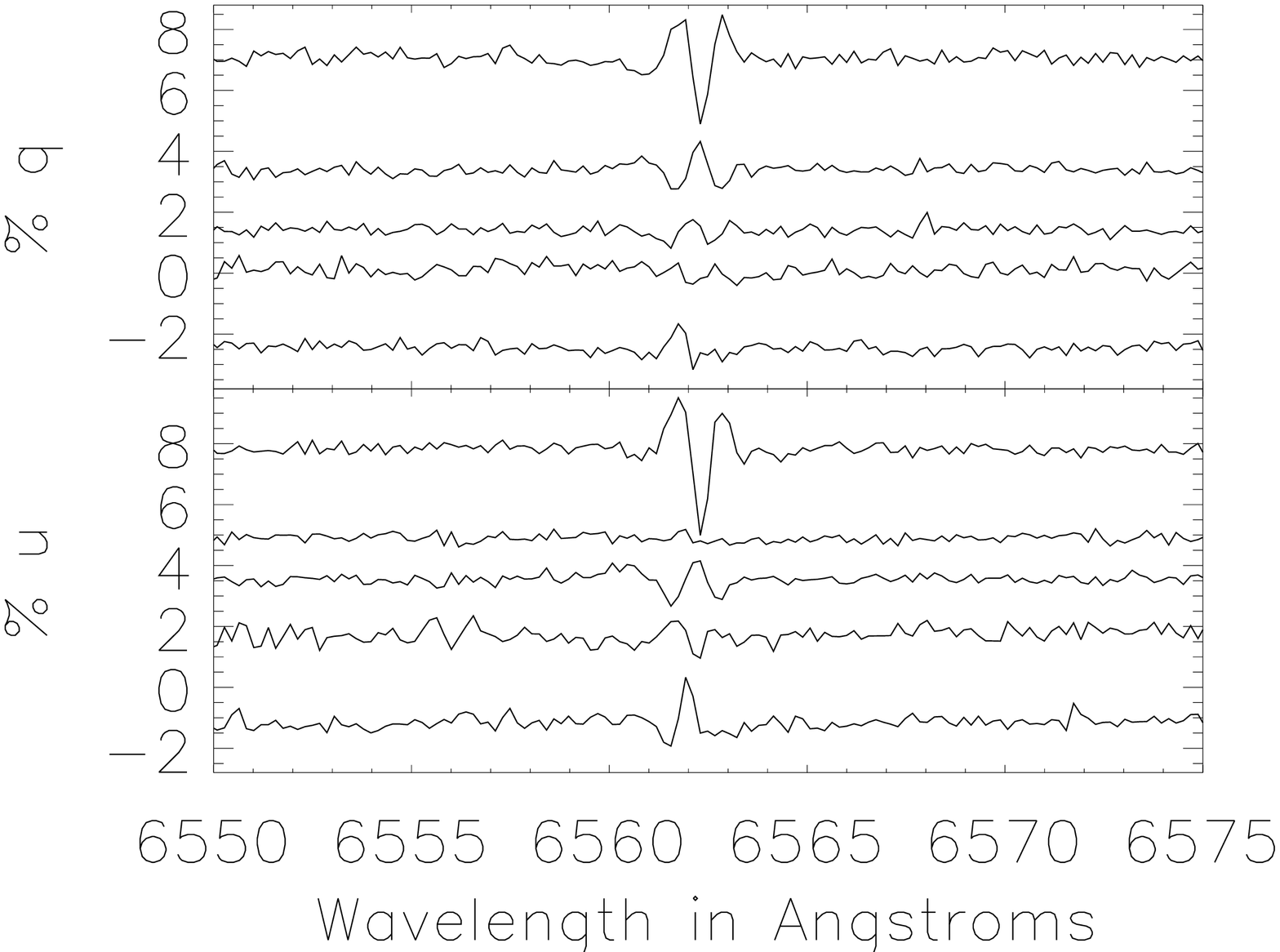}  
\caption{\label{rvtau-acher} The HiVIS H$_\alpha$ spectropolarimetry AC Her. On the top, all the q, u and intensity spectra are overplotted. The bottom box shows the unrotated q and u spectra separated vertically. The pointing varies by roughly 10$^\circ$ in azimuth and 30$^\circ$ in elevation as can be seen in the observing log of table \ref{hivisobs}. Time increases vertically. The second observation from June 27th has not been included because of possible interference from severe cirrus clouds.}
\end{center}
\end{figure}

	The three RV-Tau type variables U Mon, RU Cen and AC Her all show results broadly similar to the Post-AGB stars but with individual differences. The ESPaDOnS spectropolarimetry of U-Mon for February 2006 and HiVIS observations of 2008 are shown in figure \ref{rvtau-umon}. The top panel shows q u and intensity spectra for two ESPaDOnS observations of U Mon taken on consecutive nights. There was a strong mildly red-shifted absorptive component on top of the H$_\alpha$ emission that showed a narrow mostly antisymmetric u polarization with a somewhat broader, double-peaked q spectrum. The morphology is complex, having multiple components varying strongly over a narrow wavelength range. The bottom panel of figure \ref{rvtau-umon} shows U Mon observed with HiVIS roughly 30 months later. The H$_\alpha$ line has changed drastically, going from emission at 2.5 times continuum with a strong red-shifted absorption to just below continuum to a line barely 20\% above continuum. The spectropolarimetric morphology has also changed drastically. There is only a broad monotonic increase in q, though at a similar magnitude to the ESPaDOnS observations. 
	
\begin{figure*} [!htb]
\begin{center}
\includegraphics[width=0.23\linewidth, angle=90]{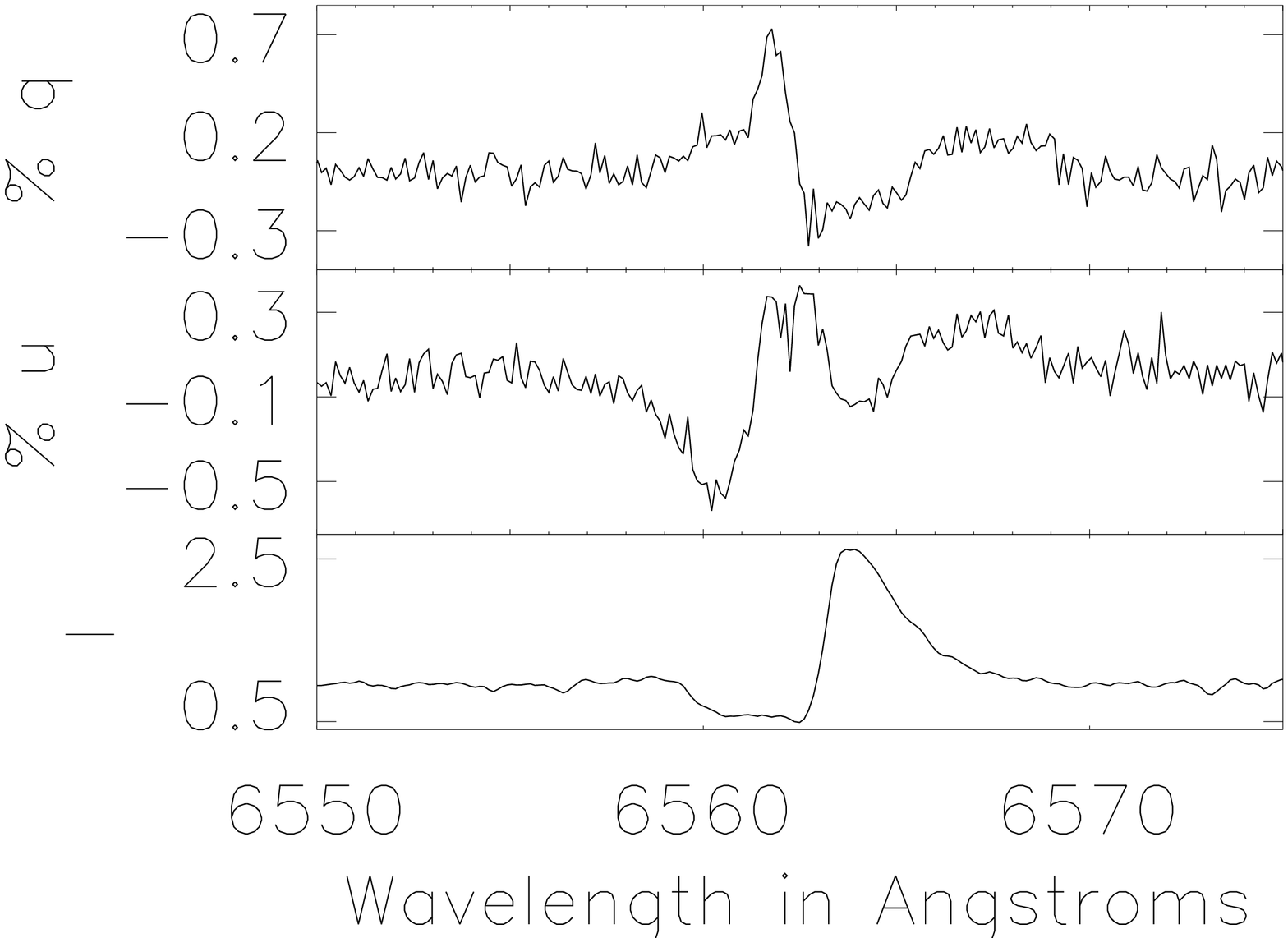}  
\includegraphics[width=0.23\linewidth, angle=90]{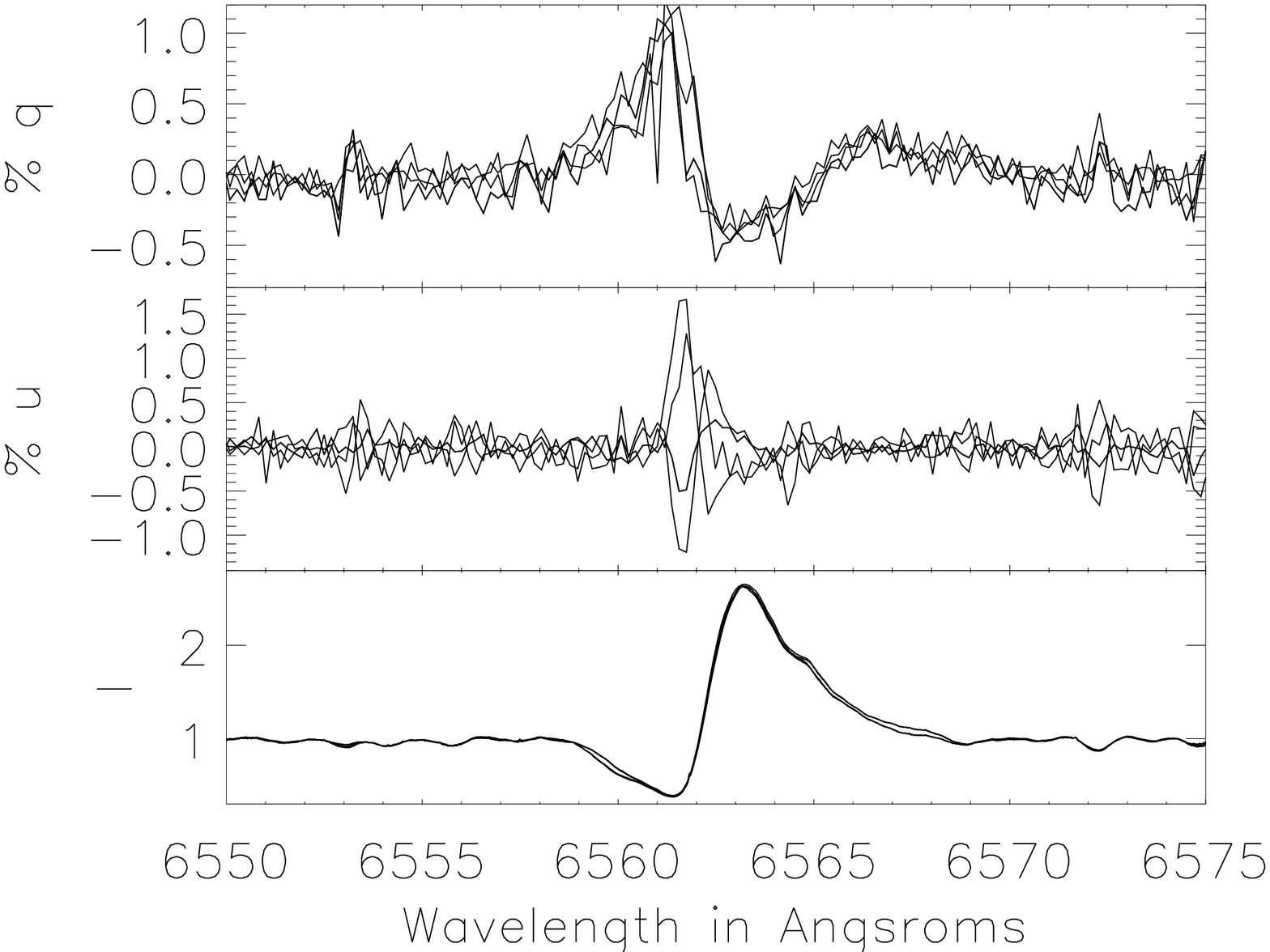}  
\includegraphics[width=0.23\linewidth, angle=90]{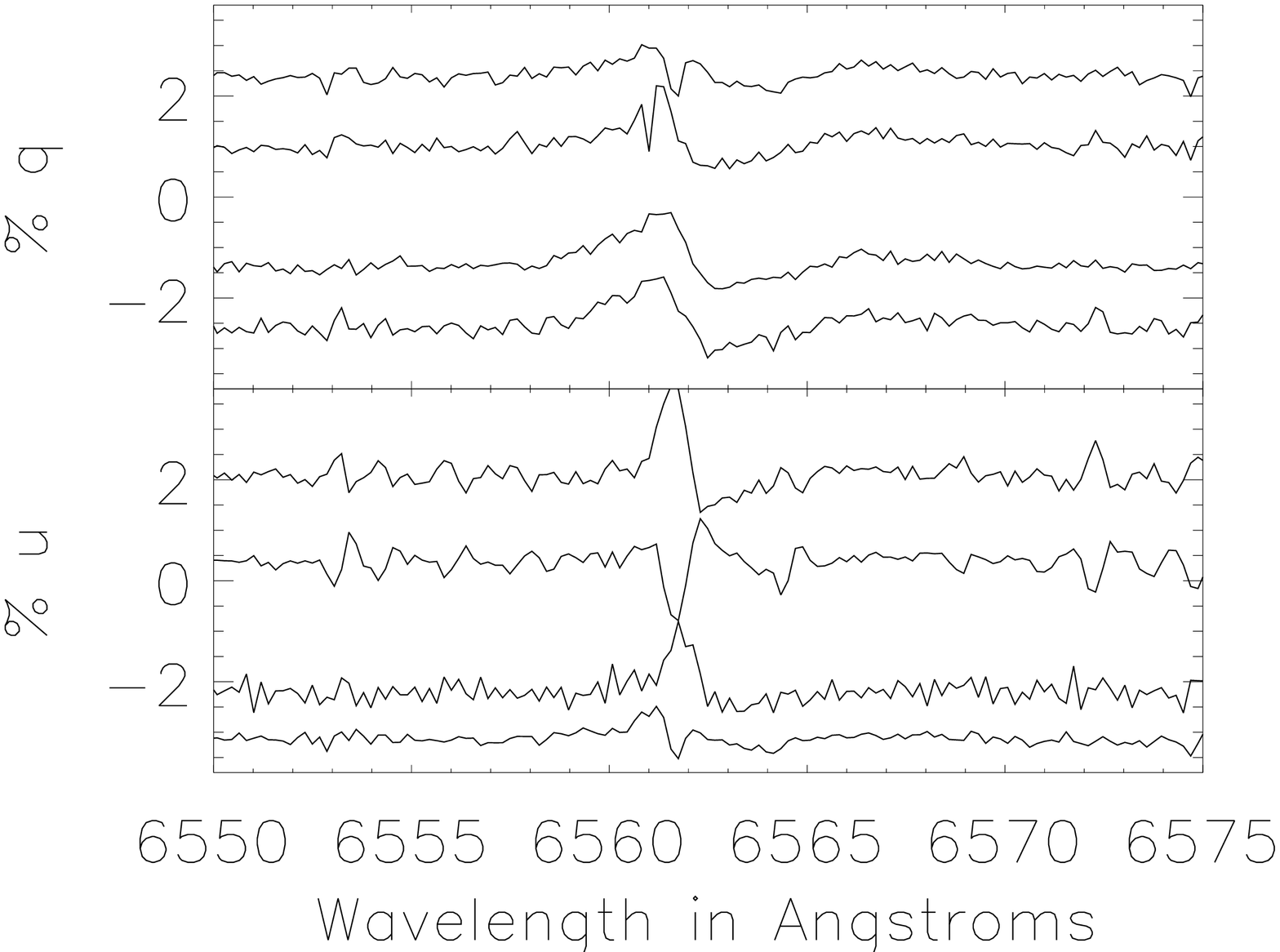}  
\caption{\label{sslep} The H$_\alpha$ spectropolarimetry for SS Lep. The first plot shows the ESPaDOnS archive observations. The second plot shows the q, u and continuum-normalized intensity spectra. The q and u spectra have been rotated to maximize the q spectrum in the blue-shifted wing (6561-6562\AA). The third plot shows the q and u spectra in the unrotated HiVIS frame. Each night has been offset for clarity. Time increases vertically. The observing log in table \ref{hivisobs} shows that these observations are on consecutive nights and are taken back-to-back. The strong change seen in the top two u spectra, from 9-28 show that this target is strongly variable on hour time-scales.}
\end{center}
\end{figure*}
		
	Figure \ref{rvtau-esp} shows the ESPaDOnS archive data for RU Cen (top) and AC Her (bottom). Both show central absorption over significant H$_\alpha$ emission with relatively complex structure on the red-shifted side. The detected polarization is relatively simple - a somewhat broad change in q and u, though with a small wavelength difference between q and u. In qu-space, these are simple qu-loops. The figure \ref{rvtau-acher} shows the HiVIS observations for AC Her taken in autumn of 2008. The second observation from June 27th has not been included because of possible interference from severe cirrus clouds. The top panel shows the q u and intensity spectra for all the observations overlaid. The bottom panel shows the observations vertically separated with time increasing towards the top. The variability in H$_\alpha$ morphology seen in the HiVIS observations of AC Her is quite significant. The observations only span one month, June 27 to July 27 2008, but the emission strength nearly doubles and the center of the absorptive component shifts toward the red. The four curves showing stronger emission also vary significantly with time, the blue-shifted emission peak going from around 1.8 to 2.2 in the span of 3 days. The q and u spectra for each observation are shown in the bottom panel in the unrotated HiVIS instrument frame. The observations were taken at significantly different pointings spanning 13$^\circ$ in azimuth and 38$^\circ$ in elevation. Though the QU frame is not consistent, the morphology of the signatures can be compared. The overall morphology of every HiVIS observation is somewhat different from the ESPaDOnS observations. There is a noticeable three-peak structure in every HiVIS set while the ESPaDOnS observations are monotonic. The amplitude varies quite significantly, going from barely-detectible to an amplitude of nearly 2\% seen in one of the observations though this could be caused by AEOS telescope polarization. The spectra in figure \ref{rvtau-acher} represent June 27th, July 24th, 2 on July 26th and finally July 27th. There is a noticeable change in the two July 26th observations with a relatively small change in pointing, suggesting short-term variability.

	The main result of the Post-AGB and RV-Tau observations is that the detection of absorptive polarimetric effects is very common, occurring in 8/9 of these stars. The polarimetric effect is entirely contained in the absorptive component of the line for almost all observations and covers a small wavelength range. 89 Her is somewhat of an exception but shows a very strong P-Cygni type absorption that almost entirely swallows the emission line. U Mon is also an exception in the HiVIS observations. The H$_\alpha$ morphology changed very significantly between the HiVIS observations showing a broad spectropolarimetric effect and the ESPaDOnS observations showing the typical polarization-in-absorption effect. Almost all targets observed were variable in time when data existed to test for variability. There is significant variability with time in the absorptive polarization morphology in most sources with multiple epochs of observation - 89 Her, AC Her and U Mon. PS Gem is quite possibly variable but more observations are needed to confirm this observation.
	
	There is no compelling evidence for any significant difference between the pulsing RV-Tau subclass spectropolarimetry and their Post-AGB counterparts. The broad morphology and detection rate is similar to the Herbig Ae/Be stars - most Post-AGB stars show this absorptive polarimetric effect as do most Herbig Ae/Be stars even though there is a very large diversity between individual stellar systems and spectropolarimetric morphologies.

\section{SS Lep} 

	Since there has been very significant variability observed in several sources now and the stars showing strongest P-Cygni type absorption seem to have the strongest polarimetric signatures and be the most variable, examination of other star types showing P-Cygni profiles seems important. SS Lep is another interesting example of an A-type star showing a P-Cygni type H$_\alpha$ line profile with remarkably different physical circumstances than other stellar types. SS Lep (HD 41511, V=5.0) is a semi-detached binary with A and M type stars with a general mass ration of 4:1 showing the Algol Paradox. The system is a main sequence A star and a giant M star filling its Roche lobe. The Algol Paradox is that the most evolved star in the the binary pair is the least massive, being caused by past mass transfer. This system has not shown evidence for eclipses so it is not a ``true'' Algol star (cf. Verhoelst et al. 2007). Welty \& Wade 1995 present a spectroscopic program that finds an eccentricity of 0.132 and a mass ratio of 3.6$\pm$0.6. A recent study (Jura et al. 2001) found the far-infrared/mm flux to come from a circumbinary disk. There is evidence for mass transfer from the shell spectrum with significant UV activity. Some have argued that this system can be modeled as a pre-main-sequence binary (Polidan \& Shore 1993). However, this was debated on several lines in Welty \& Wade 1995. Recent interferometry also estimates both the star and Roche lobe sizes at 100R$_\odot$ at the distance of $\sim$300pc and an inclination of 30$^\circ \pm$10$^\circ$ (Verhoelst et al. 2007). The study also found that half the K-band flux was from an extended circumbinary structure.

	As a consequence of the mass transfer, the star should be accreting. Verhoelst et al. 2007 used interferrometry to find a stellar radius of 18R$_\odot$ and put the star at 2M$_\odot$. This radius is an order of magnitude larger than a normal A1V star. The star was was argued to be accreting at a rate of 2 $\times$ 10$^{-4}$M$_\odot$/yr having already accreted 1 M$_\odot$ by using the evolutionary tracks for accreting main sequence stars of Kippenhahn \& Meyer-Hofmeister 1977.  The main purpose of including this star in our investigation is that the A-type star produces a clear P-Cygni type H$_\alpha$ line. This indicates mass loss from that A-star and a potentially complex circumstellar geometry close to the surface of the star.
		
	ESPaDOnS observations from February 9th 2006 binned as previously described to a regular wavelength array are shown in figure \ref{sslep}. The Libre-Esprit estimate of the signal-to-noise was around 1100 for both spectral orders used in the binning. The figure also shows HiVIS spectropolarimetry from September 2008 both rotated and unrotated. The observing log for the HiVIS observations is shown in table \ref{hivisobs}. The HiVIS spectropolarimetric frame has been rotated in the second panel of figure \ref{sslep} for each of the four observations to maximize Stokes q from 6560-6561\AA. This rotation has the effect of making the ESPaDOnS and HiVIS q spectra appear morphologically similar. This rotation highlights any variability. Though the q spectra are roughly similar in all the HiVIS observations, the u spectra differ quite strongly and show significant short-term variability. The four HiVIS observations were taken in two sets of two observations on consecutive days with only a small difference of roughly 5$^\circ$ between azimuths and elevations. Though observations are back-to-back on consecutive nights, there is very significant variability seen on both nights between the consecutive observations. This suggests strong short-timescale variability. The third plot of figure \ref{sslep} shows the q and u observations in the HiVIS unrotated frame with time increasing vertically. In particular, the top two u spectra (from September 28th) show how the strong u absorptive polarization signature switched sign over roughly one hour. On the previous night, September 27th, there was a similar switch though with much lower amplitude. On all nights, the q spectra are all mostly similar, showing mostly the morphology of the rotated q spectra.

\section{Comparison to Herbig Ae/Be Stars}

	There are many similarities between the spectropolarimetric morphologies of the Post-AGB stars (with the associated RV-Tau type variables) and the Herbig Ae/Be stars presented in Harrington \& Kuhn 2009. In this section we make two main points - 1) the absorptive polarization effect is similar in morphology and strong time variability between Post-AGB and Herbig Ae/Be stars and 2) a member of the pulsing subclass of Herbig Ae stars (the $\delta$ Scuti type) also shows no significant difference in it's polarimetric behavior to the broad Herbig Ae class. 

\subsection{Absorptive Variability}
	
	The strong absorptive polarization variability seen in 89 Her, AC Her and SS Lep is fairly typical of the Herbig Ae/Be stars as well. We present here a few example ESPaDOnS and HiVIS observations to support this case. Since the HiVIS polarimetric frame and polarization properties are pointing-dependent, only ESPaDOnS observations can be used to simultaneously measure both amplitude and morphological changes over both short and long time-scales. HiVIS observations may be used but care must be taken to make sure the observations are at similar telescope pointings (to ensure similar telescope polarization properties). There are now observations of significant variability in several Herbig Ae/Be stars. Figure \ref{haebe} shows observations of AB Aurigae and MWC 120.  Both AB Aurigae and MWC 120 were observed in February 2006 and August 2008 and show significant changes in the absorptive spectropolarimetric effect. The AB Aurigae signature greatly increased strength in the P-Cygni trough though the absorption strength in intensity was roughly the same at 0.5 times continuum. MWC 120 showed it's magnitude drop by more than a factor of two in the central narrow absorptive component though the difference in intensity at these wavelengths shows only a small change. This is quite similar to the variability presented for MWC 480 in Harrington \& Kuhn 2009. There are also stars where the spectropolarimetry does not change much, such as MWC 361, even though the overall line intensity and absorption profiles do.

\begin{figure} [!h]
\begin{center}
\includegraphics[width=0.6\linewidth, angle=90]{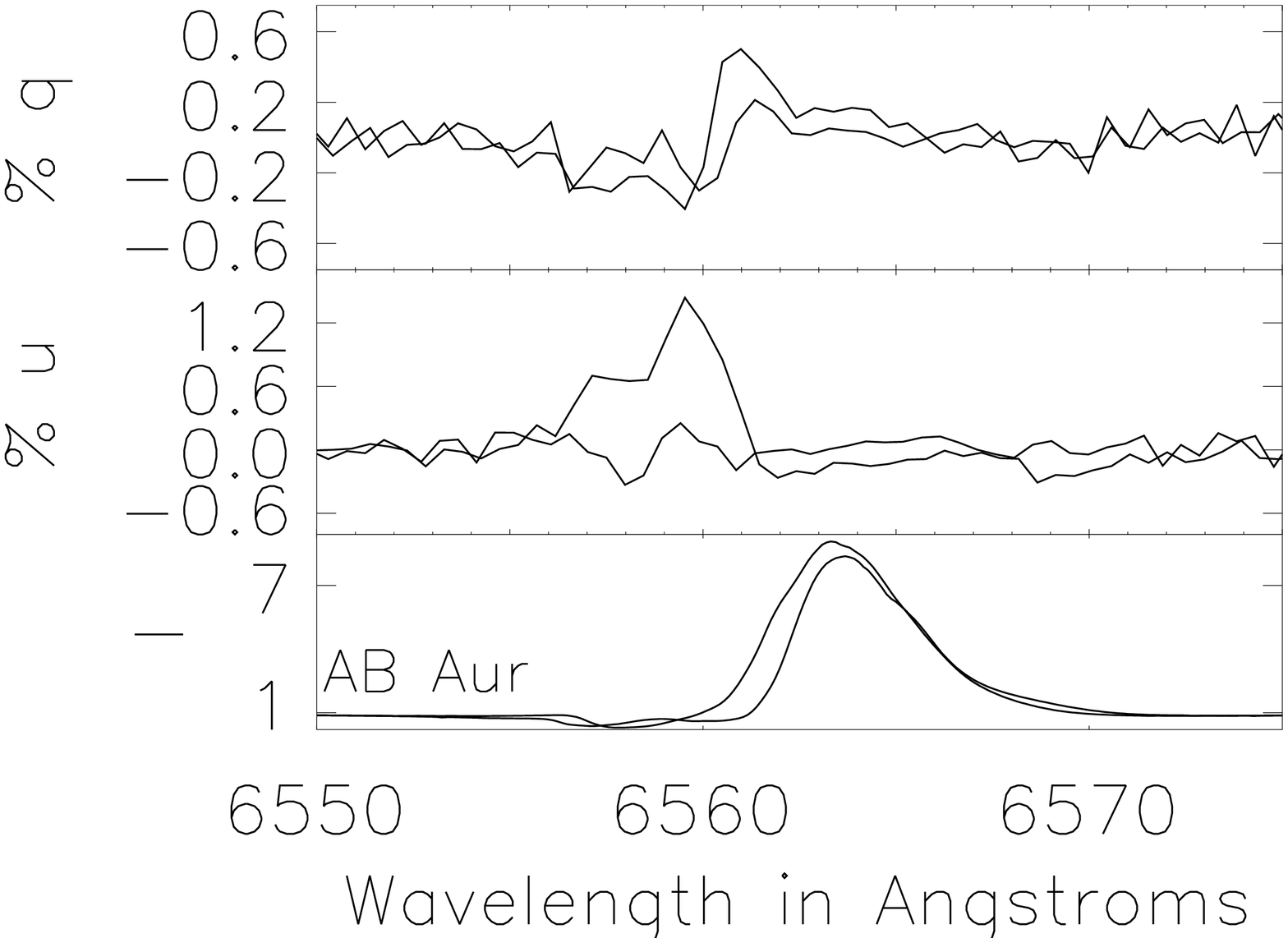}  
\includegraphics[width=0.6\linewidth, angle=90]{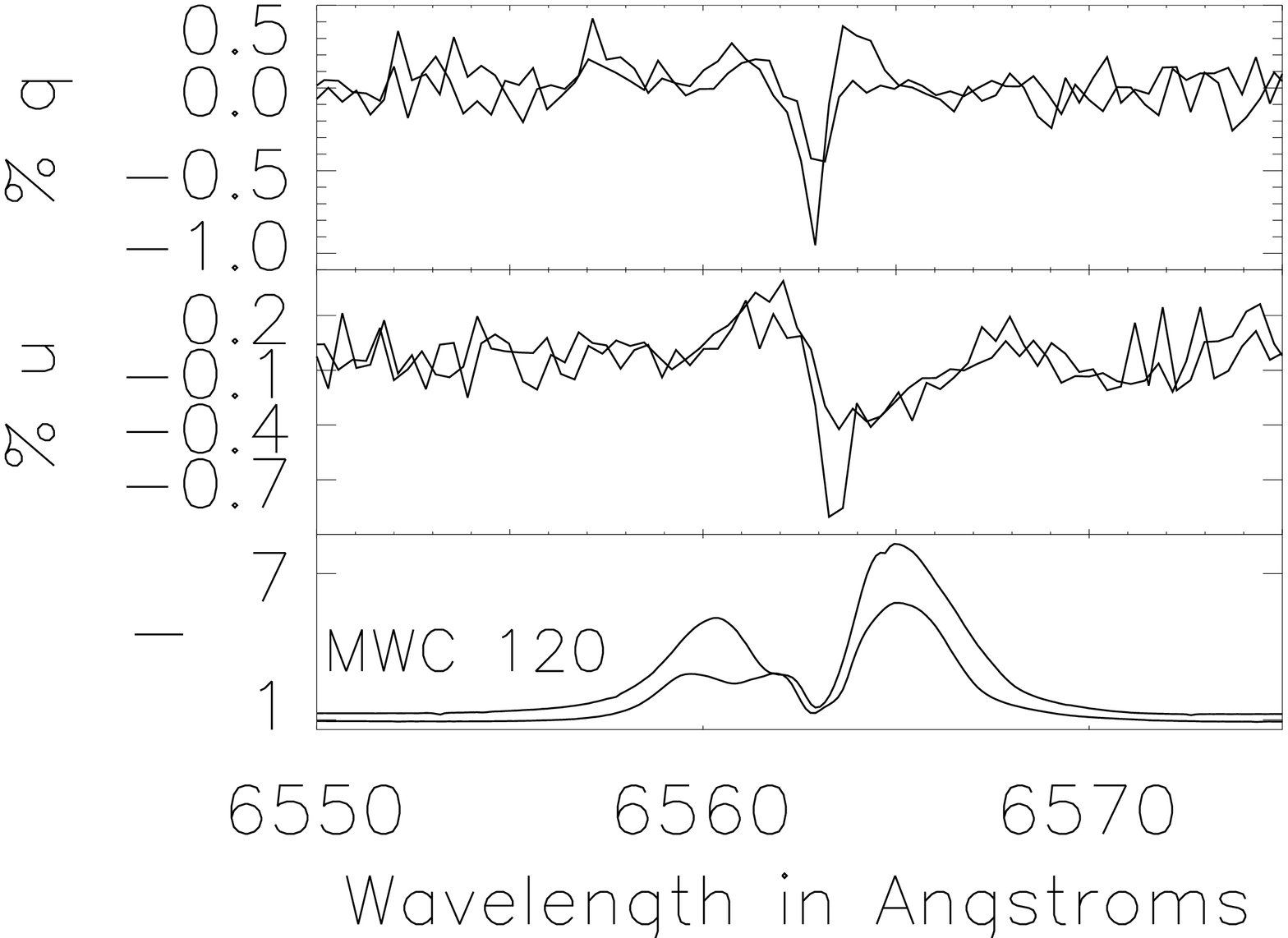}  
\caption{\label{haebevar} ESPaDOnS H$_\alpha$ spectropolarimetry for the Herbig stars AB Aurigae and MWC 120. The two AB Aurigae observations shown in the top panel were taken February 7th 2006 and August 23rd 2008. MWC 120 is shown in the bottom panel and was observed February 9th 2006 and August 25th 2008. Both observations have been binned by an additional factor of 4 to clearly show the change in spectropolarimetric magnitude and morphology. AB Aurigae has a signature that greatly increased in u magnitude from 2006 to 2008. The MWC 120 signature decreased by more than a factor of 2 in the same time period.}
\end{center}
\end{figure}

	As an example of the Herbig Ae stellar class, we will describe with the largest number of observations and the strongest absorptive polarization variability, AB Aurigae. This star has a near face-on circumstellar disk resolved in many wavelengths (eg: Grady et al. 2005, Fukagawa et al. 2004). It also has an active stellar wind with it's strong emission lines often showing strong P-Cygni profiles. Spectroscopic measurements put AB Aurigae somewhere between late B and early A spectral types (B9 in Th\'{e} et al. 1994, B9Ve in Beskrovnaya et al. 1995, A0 to A1 Fernandez et al. 1995) with an effective photospheric temperature of around 10000K. 
		
	The star has a wind that is observably non-spherically symmetric with a mass loss rate of order $10^{-8} {M_\odot}$ per year. Spectroscopic models favor an extended chromosphere reaching $T_{max}\sim$17000K  around 1.5$R_\ast$ (Catala \& Kunasz 1987, Catala et al. 1999). Bouret et al. 1997 found N V in the wind which require clumps of T$\sim$140,000K material at a filling factor of $\sim10^{-3}$. 

	AB Aurigae shows spectropolarimetric variability in our HiVIS observations on day-to-day time-scales, similar to SS Lep and 89 Her. The compiled results of the large HiVIS observing campaign were presented in Harrington \& Kuhn 2007. Over 160 individual spectropolarimetric observations have been collected in our survey so far from 2006 to 2008. Spectroscopically, there is small-amplitude but significant variation in the H$_\alpha$ intensity over 10-minute time scales. A short-term variability study done by Catala et al. (1999) showed that an equatorial wind with a variable opening angle, or a disk-wind originating 1.6$R_\ast$ out with a similar opening angle could explain the variability. Bohm \& Catala (1993, 1995) present a complete spectral atlas and followup work on stellar activity. Bohm et al. 1996 examined changes in the P-Cygni absorption in order to derive a stellar rotation period of 32 hours. A detailed spectroscopic examination of AB Aurigae was presented in Bouret \& Catala 1998.
	
\begin{figure} [!h]
\begin{center}
\includegraphics[width=0.6\linewidth, angle=90]{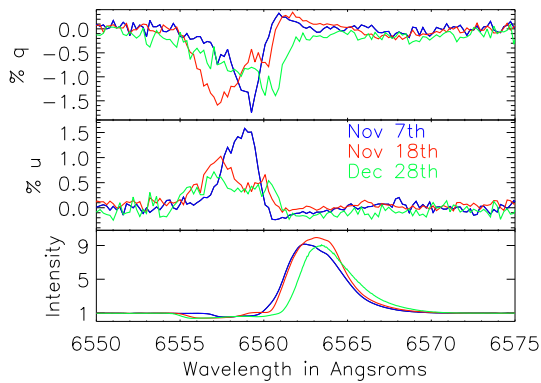}  
\caption{\label{abvar} HiVIS spectropolarimetric observations of AB Aurigae showing the temporal variability of the spectropolarimetric signatures. Each curve corresponds to an average of four to seven individual data sets taken at 'rising' pointings. See text for details. There is significant variability in width, relative amplitude and morphology between all three observations.}
\end{center}
\end{figure}
	
	There is a fortunate combination of pointing and AEOS polarization properties for this star enabling a comparison of observations over a wide range of time-scales. When rising and setting, AB Aurigae essentially changes only in altitude at two fixed azimuths (60 and 300) due to the similarity in declination and Hawai'i latitude. The azimuth change rates only become significant when at altitudes above 70 for our typical exposure times of 8-30 minutes per spectropolarimetric set. The polarization properties of the telescope are fairly benign. AB Aurigae consistently showed a (-Q +U) signature when `rising' and a (+Q +U) signature when `setting'. There was not much systematic variation seen in amplitude or morphology in many nights of observations when pointing between altitudes of 30 to 75. On the several nights when we tracked the target through transit, the spectropolarimetric signature decreased in amplitude and Q reversed sign, consistent with rotational smear (with azimuth) and rapidly changing telescope properties. When setting, the spectropolarimetric signatures were similar but with a reversed sign for q. This means that the magnitudes and morphologies observed are easily comparable when taken at these telescope pointings.

	The spectropolarimetric variability between three nights is shown in figure \ref{abvar}. During the winter 2006-2007 season we monitored this star quite heavily on several nights. The figure shows November 7, 18 and December 27 as examples of the night to night variation typically seen in the observations. Each curve here represents an average of 4 to 7 individual spectropolarimetric observations taken as AB Aurigae was rising. The short-term variability between individual exposures used in the averaging was very small on these particular nights, being indistinguishable from the noise.

	The observations show distinct differences between the central wavelengths of the q and u spectra. The u spectra all peak around the same wavelengths while the q spectra vary significantly. The relative magnitudes also vary. The u spectra from November 7th peaks around 1.5\% while the other two nights show significantly less than this while the q spectra are all roughly similar amplitudes. The overall absorptive profiles vary significantly in morphology between the nights but the depth of the P-Cygni absorption is consistently 0.4 to 0.5 times continuum. The width of the detected spectropolarimetric effect varies significantly and it correlates with the P-Cygni absorption width, as was outlined in Harrington \& Kuhn 2007.

	The strong and sometimes very short-term variability seen in Post-AGB (89 Her) and Herbig Ae/Be (AB Aur) as well as other types (SS Lep) argues for the universal and close-in nature of material causing the polarization in every case. It also points out the need of any survey project to have multiple epoch observations. Spectropolarimetric signatures can appear and disappear over day to year periods, as seen in several targets presented here.

\begin{figure*} [!ht]
\includegraphics[width=0.23\linewidth, angle=90]{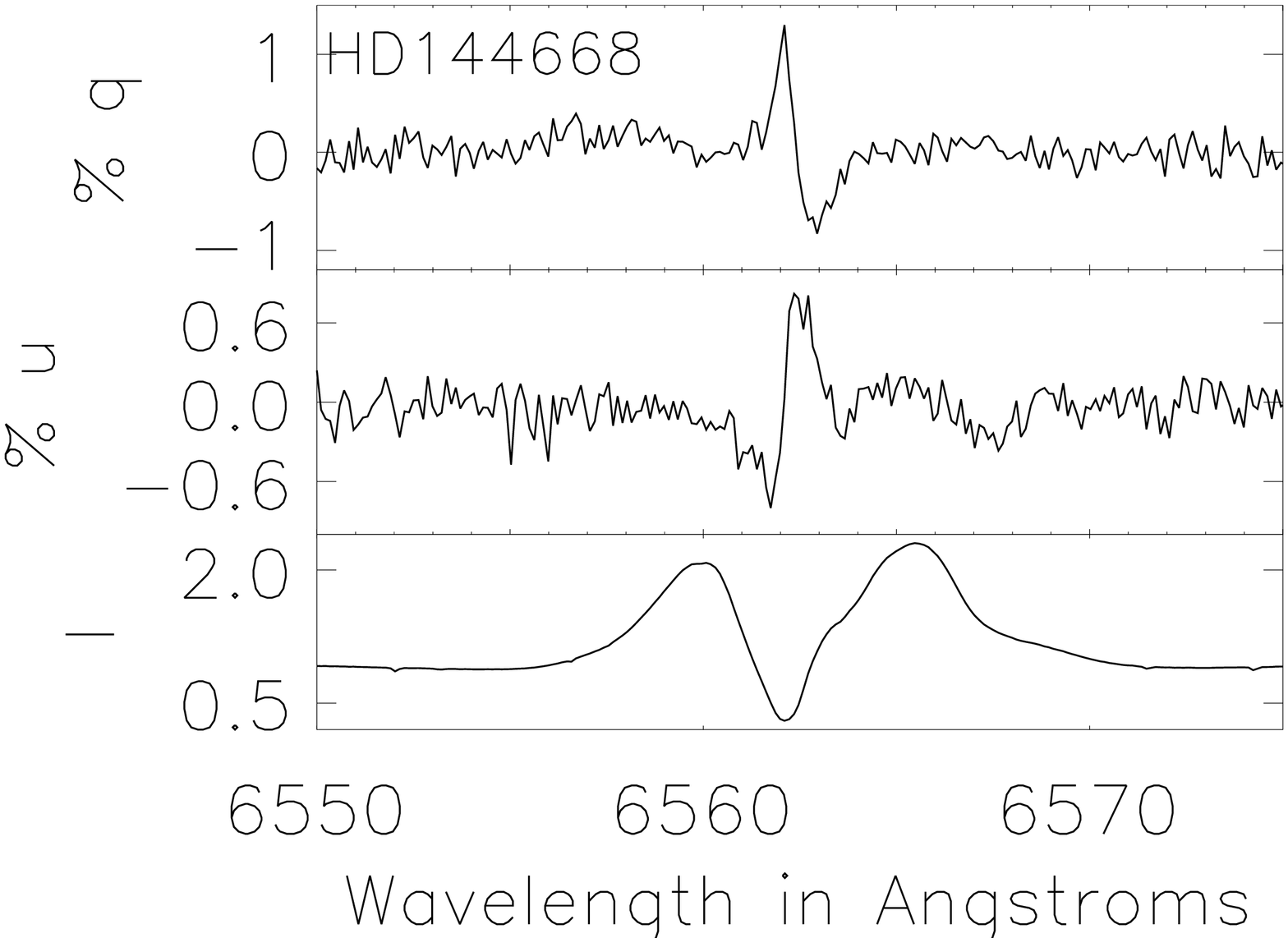}  
\includegraphics[width=0.23\linewidth, angle=90]{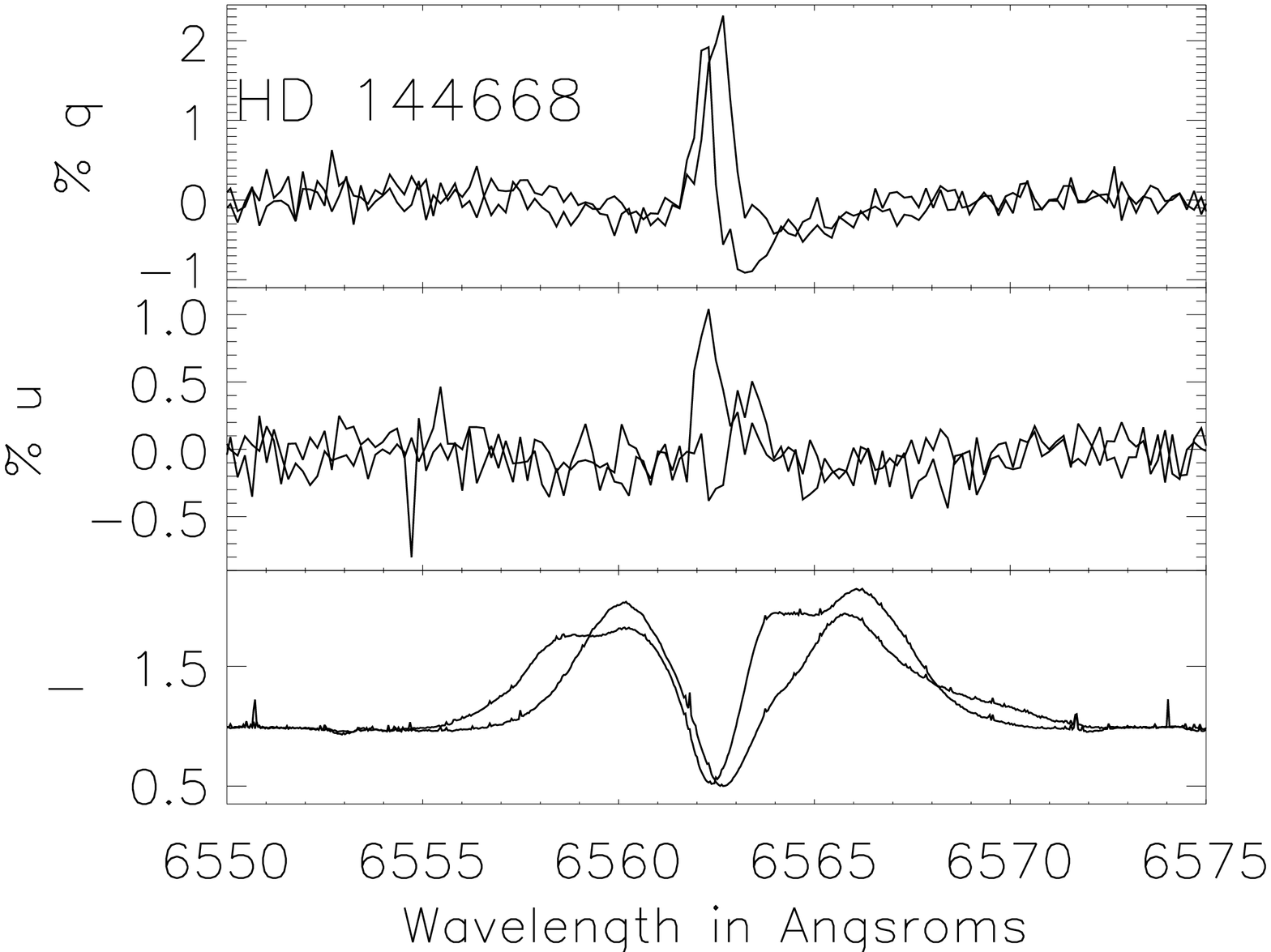}  
\includegraphics[width=0.23\linewidth, angle=90]{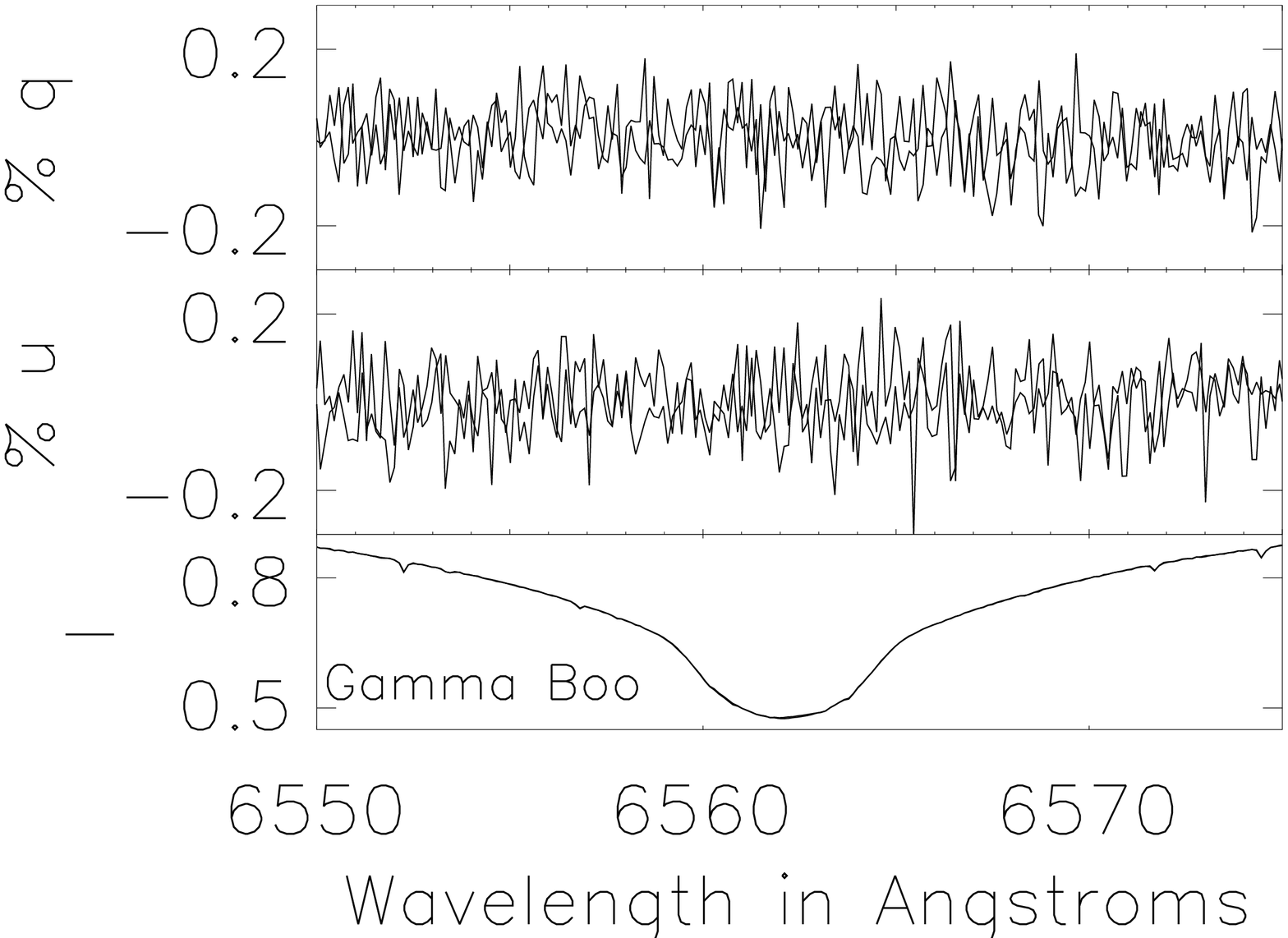} 
\caption{\label{haebe} This shows the H$_\alpha$ spectropolarimetry for the Herbig Ae star HD 144668, an A7 IVe $\delta$-Scuti variable in the Simbad database and $\gamma$ Boo, an A7III $\delta$-Scuti variable. The left panel shows ESPaDOnS archive spectropolarimetry of HD 144668 from August 14th 2006 with a signal-to-noise of 569. The middle panel shows two HiVIS observations of HD 144668 in the unrotated HiVIS frame from June 27th and July 27th 2008 (173, 31) and (189,30) az-el. The right panel shows ESPaDOnS archival observations of $\gamma$ Boo from June 10th and 11th 2006.}
\end{figure*}
	
\subsection{Pulsating Herbig Ae stars - $\delta$-Scuti sub-type}	

	The Herbig Ae stars have a sub-class of pulsing stars, the $\delta$-Scuti type variables. $\delta$ Scuti variables are intermediate-mass stars, 1.4 - 3.0M$_\odot$ with A to F spectral type in the lower part of the classical instability strip. They have small amplitude variations, typically about 0.02 magnitudes up to several tenths (cf. Rodr\'{i}guez et al. 2000). 
	
 	Three of the Herbig Ae stars in Harrington \& Kuhn 2009 are included in the Zwintz 2008 list of $\delta$-Scuti type variables with detected oscillations: HD 144668, HD 142666 and HD 35929. The star HD 144668 (HR 5999, V856 Sco) shows a broad H$_\alpha$ line with a strong central absorptive feature and has a strong absorptive polarization effect. This star showed 13 mmag amplitude pulsation with a 5.0 hr period in Johnson V with a background variability of 350 mmag assumed to come from circumstellar material (Kurtz \& Marang 1995). In both ESPaDOnS and HiVIS spectropolarimetry, shown in figure \ref{haebe}, the polarization effect is narrow and entirely contained in the central absorptive component. In the two years between observations, the antisymmetry of the HiVIS observations are somewhat reduced and the magnitude nearly doubled. The overall character though remains unchanged.
	
	The other two stars, HD 142666 and HD 35929 are non-detections in the observations presented in Harrington \& Kuhn 2009 (figures 5 and 24). HD 142666 showed a mildly red-shifted absorption component on a broad line of roughly 2 times continuum. This star's H$_\alpha$ line looked quite similar to the ESPaDOnS observations of U Mon, though the polarimetric sensitivity in Harrington \& Kuhn 2009 was only 0.5\%.  HD 35929 did not show evidence of strong absorption and was a non-detection at the 0.3\% level. It should be noted that Miroshnichenko et al. 2004 concluded that this star was an F2 III type star at 2.3M$_\odot$ and agreed with Marconi et al. 2000 that this star is not a young object but instead a post-main-sequence giant in the instability strip with significant mass loss. Though there are observations of only a few stars in this sub-class, there appears to be no substantive difference between these Herbig Ae $\delta$-Scuti variable stars and the broader Herbig Ae class. 
	
	As an example, figure \ref{haebe} also shows spectropolarimetry of a main-sequence classical $\delta$-Scuti type star, $\gamma$ Boo with no detectable effect above the 0.1\% level. This star has a simple H$_\alpha$ profile showing no obvious evidence for circumstellar emission or absorption. This non-detection is also consistent with all the non-detections for the unpolarized standard stars showing simple H$_\alpha$ profiles presented in Harrington \& Kuhn 2007. No spectropolarimetric effect has ever been found in stars showing similar H$_\alpha$ profiles like $\gamma$ Boo.

\section{Discussion}

	Spectropolarimetric signatures have been detected in Post-AGB stars at a very high frequency - 8/9. The magnitude of the detected spectropolarimetric effects go from 0.5\% up to over 3\% and is quite similar to the polarization-in-absorption seen in previous studies. There very significant absorptive spectropolarimetric variability in 89 Her, AC Her and U Mon with possible variability in PS Gem. This is similar to the variability seen in the Herbig Ae/Be stars (AB Aurigae, MWC 120 and MWC 480) as well as SS Lep, an A-type star showing a strong P-Cygni type H$_\alpha$ profile. There is no evidence that stellar pulsations (RV-Tau types for Post-AGB stars or $\delta$-Scuti types for Herbig Ae stars) have any influence on spectropolarimetric morphology. The short (hour to week) time-scale variability of the polarized spectra of many of our target stars is consistent with a near-stellar location for the absorbing/polarizing gas -- as the optical pumping model requires. 
		
	The Herbig Ae/Be stars shown in Harrington \& Kuhn 2009 did not fit any of the typical scattering morphologies and were dominated by absorptive polarimetric effects. The disk-scattering models produce double-peaked spectropolarimetric morphologies that must be broader than the entire H$_\alpha$ emission line, contrary to what we observe. The classical depolarization signature produces a smooth broad line profile change due to unpolarized emission which dilutes the continuum polarization. The continuum is polarized by scattering in a flattened circumstellar envelope. This depolarization signature can be modified by absorption - circumstellar material outside the envelope and emission region can selectively absorb unpolarized stellar light or polarized envelope light modifying the depolarization effect. In this case, the circumstellar material must be outside the H$_\alpha$ formation region and it must act on an effect that is formed over the entire line.

\begin{figure*} [!h]
\begin{centering}
\includegraphics[width=0.35\linewidth, angle=90]{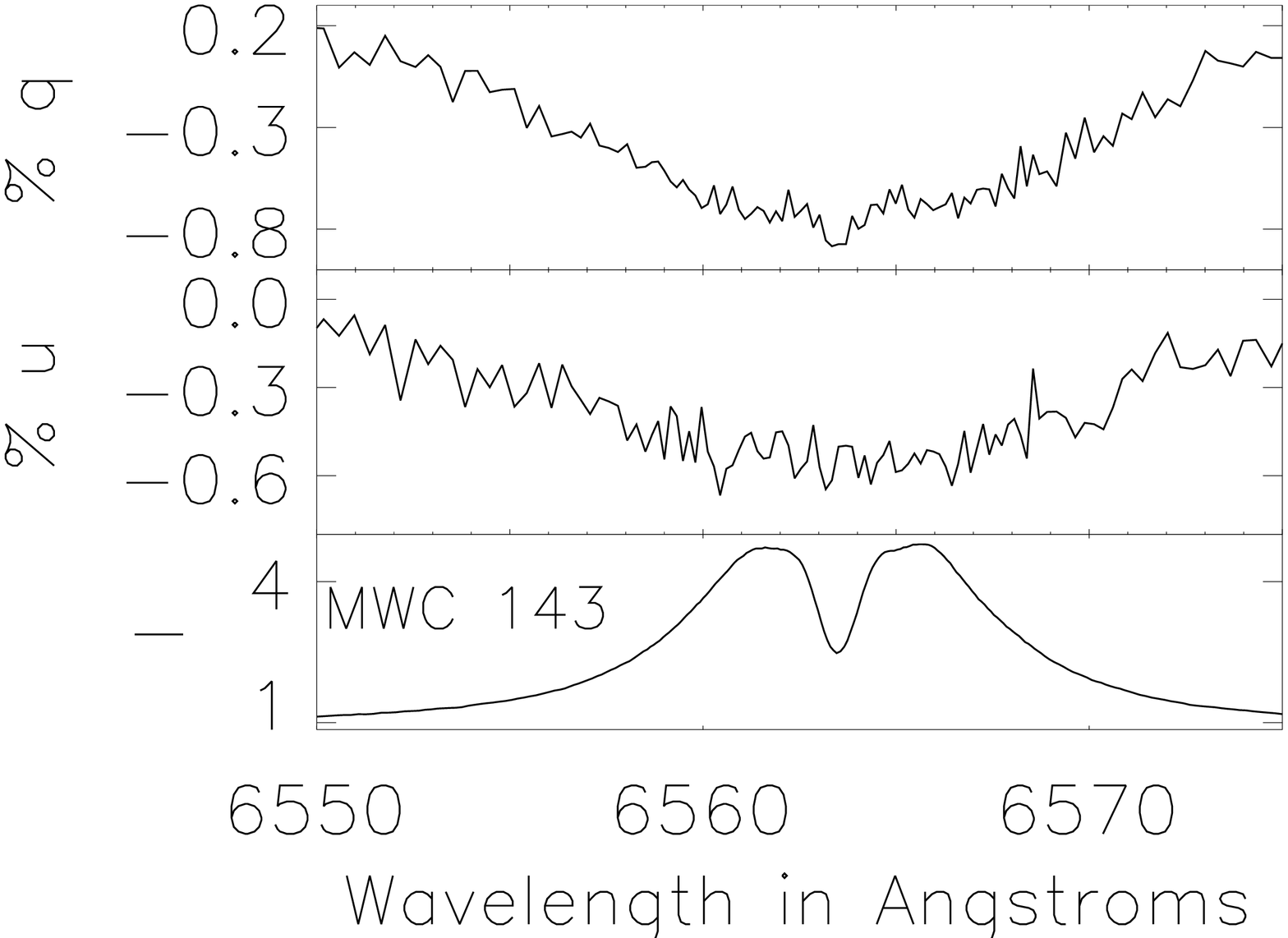}  
\includegraphics[width=0.35\linewidth, angle=90]{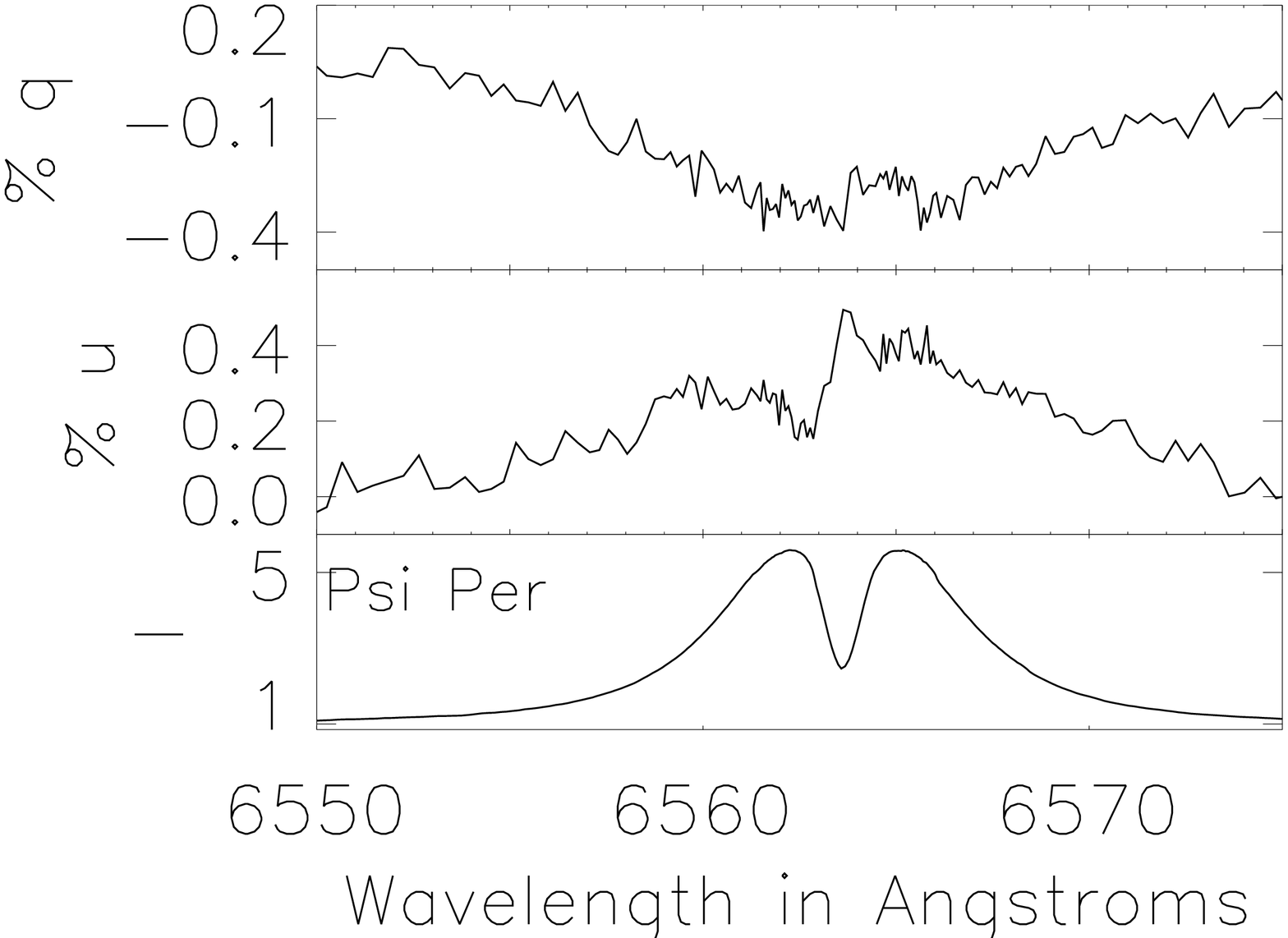}  \\ 
\includegraphics[width=0.35\linewidth, angle=90]{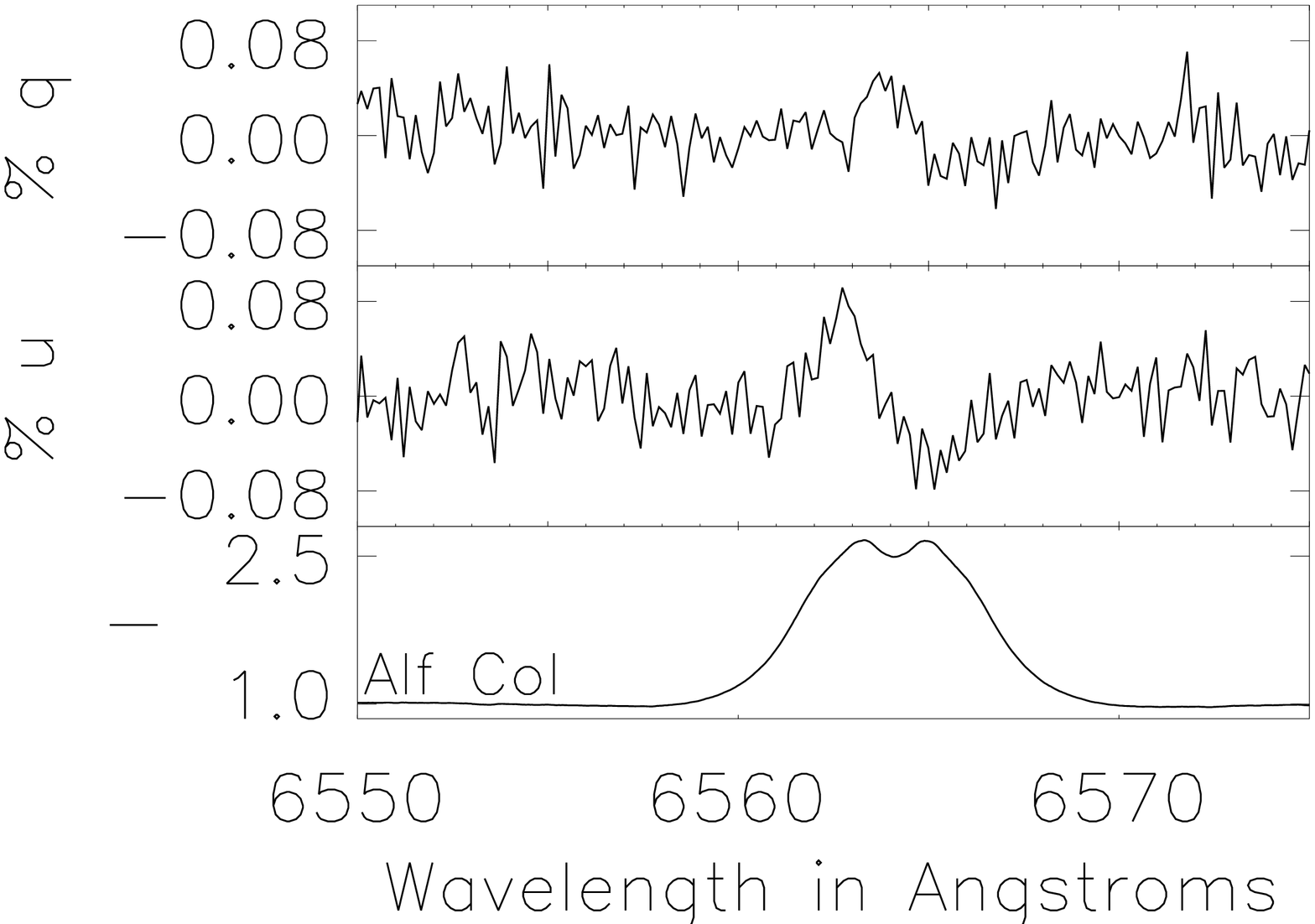}  
\includegraphics[width=0.35\linewidth, angle=90]{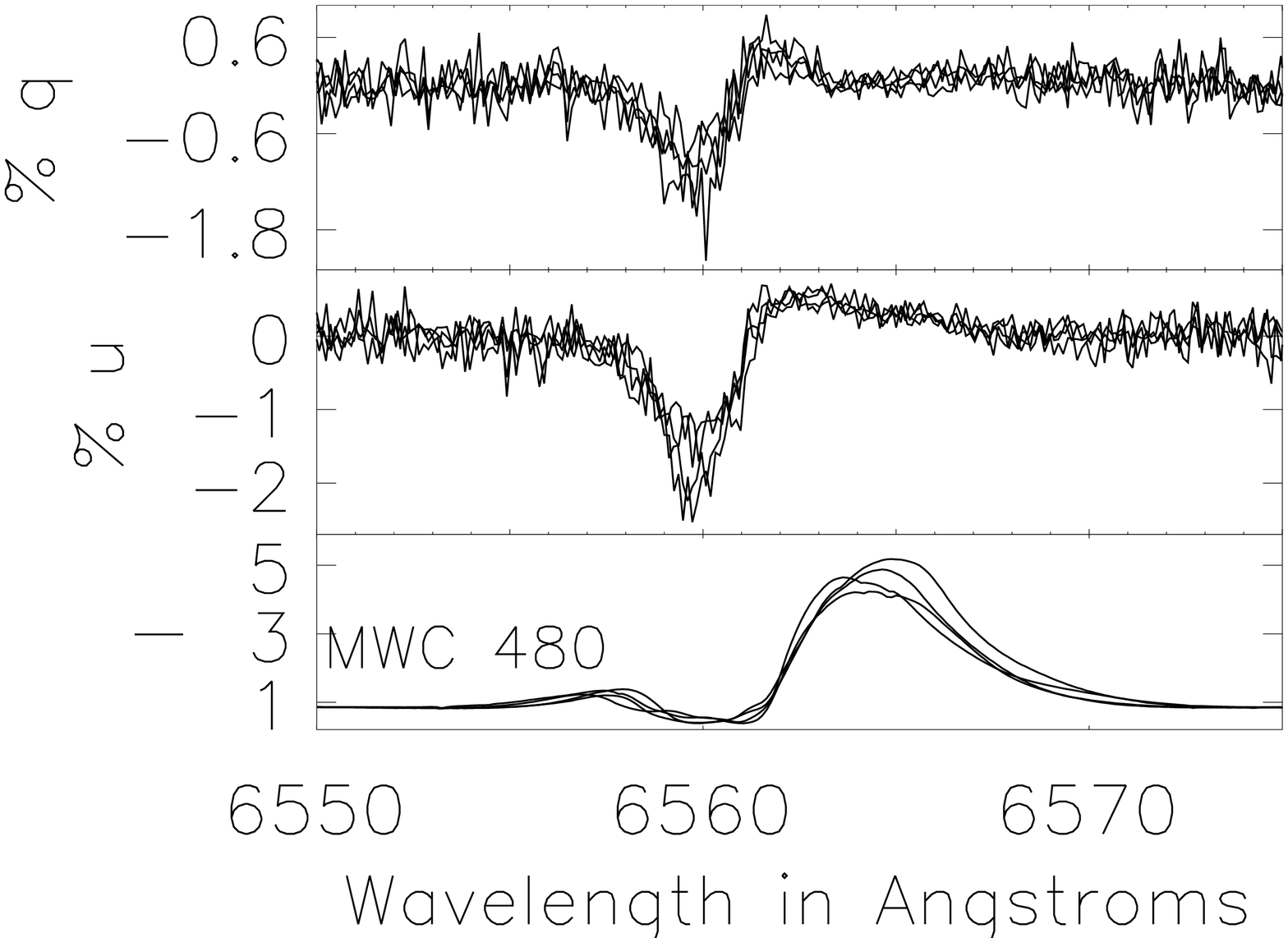} \\
\caption{\label{scomp} Comparison spectropolarimetry with Herbig Ae/Be and other B-type stars. Spectropolarimetry taken with the HiVIS spectropolarimeter for the Be stars MWC 143 and $\psi$ Per are shown in the top left and right panels. In both stars there are strong central absorptive components but a strong absorptive effect is only seen in $\psi$ Per with a small deviation from the broad effect present in MWC 143. HiVIS spectropolarimetry for the Be star $\alpha$ Col is seen in the bottom left. Though there is an obvious central absorptive component to the H$_\alpha$ line, the spectropolarimetric signature spans most of the line width. ESPaDOnS H$_\alpha$ spectropolarimetry for the Herbig Ae star MWC 480 binned to a regular wavelength grid and binned by an additional factor of four is shown in the bottoom right panel. The panel shows three independent measurements. A very clear polarimetric signature is seen in and around the absorptive components of the H$_\alpha$ line. This is entirely typical of the Herbig Ae/Be polarization-in-absorption detections. }
\end{centering}
\end{figure*}

	Figure \ref{scomp} shows examples of spectropolarimetric morphologies for other Herbig Ae/Be stars and Be/Emission-Line stars. In Harrington \& Kuhn 2009, all but one of the Herbig Ae/Be stars showed a detectable signature that correlates with the absorptive components of the line. Most effects were entirely contained inside absorptive components of the line. Roughly 2/3 of the HAe/Be stars with 'windy' or 'disky' H$_\alpha$ profiles (strongly blue-shifted or central absorptive components respectively) showed signatures of similar magnitude and morphology as these Post-AGB stars. The Be stars typically (10/30) showed a broad depolarization type signature, as shown in figure \ref{scomp} for MWC 143. However, 4 of these 10 detections showed more complex additional effects correlating with absorptive components of the H$_\alpha$ line (e.g.   $\psi$ Per in figure \ref{scomp}). A smaller sub-set of B-type stars (5/30) showed smaller amplitude and more complex morphologies like $\alpha$ Col. This star had a lower amplitude signature that was entirely stable over two consecutive nights (at two identical telescope pointings). 
	
	In the optical pumping theory of Kuhn et al. 2007, the absorptive polarization is greatest where the obscuring material is at a large star-cloud-telescope angle. In the classical P-Cygni profile, this occurs in the red-most region of the blue-shifted absorptive component. In the case of a thin circumstellar disk, this occurs at the extreme red and blue shifts of the central absorptive component of the line. In the very diverse group of stars presented here with P-Cygni profiles (AG Ant, 89 Her, SS Lep and AB Aurigae) the spectropolarimetric effects are clearly blueshifted and vary strongly with time. The effects are correlated with the presence of absorption as expected in this model.
	
	There are no detailed spectropolarimetric models which yet account for the many phenomena thought to be present in circumstellar environments. However, simple morphological arguments can illuminate the general physical circumstellar properties. The observations presented here clearly show a remarkable similarity between the spectropolarimetric morphologies of Post-AGB and Herbig Ae/Be stars. The optical pumping model naturally reproduces the, apparently quite common, absorptive spectropolarimetric signatures (8/9 P-AGB). In that model, this signal originates from material along the line of sight to the star within 1 or 2 stellar radii of its photosphere (c.f. Kuhn et al. 2007). Though the model did not include non-photospheric emission sources, the geometry can be easily scaled up to a cloud obscuring an extended emission region with a similar scaling of the region producing the spectropolarimetric effect. This is consistent with the short time-scale variability seen in these targets. Post-AGB type stars are apparently similar to Herbig Ae/Be in that they show similar wavelength dependent absorptive polarization. The absorptive quality of the dominant linear polarization makes it unlikely that scattering models can account for this spectropolarimetric morphology. These observations extend our detection of the absorptive polarization spectropolarimetric effect to another class of stars and provide further evidence that line-of-sight absorption occurring near a star causes a linear spectropolarimetric signal at absorptive line profile wavelengths.

\section{Acknowledgements}

	This program was partially supported by the NSF AST-0123390 grant, the University of Hawaii and the AirForce Research Labs (AFRL). This research used the facilities of the Canadian Astronomy Data Centre operated by the National Research Council of Canada with the support of the Canadian Space Agency. This archive provided the ESPaDOnS observations. These observations were reduced with the dedicated software package Libre-Esprit made available by J. -F. Donati. The Simbad data base operated by CDS, Strasbourg, France was very useful for compiling stellar properties.

\end{document}